\documentclass[11pt]{JHEP3}

\newcommand{\N}{{\scriptscriptstyle N}}

\newcommand{\half}{{{\textstyle\frac{1}{2}}}}
\newcommand{\quarter}{{{\textstyle\frac{1}{4}}}}

\newcommand{\be}{\begin{equation} }
\newcommand{\ee}{\end{equation} }
\newcommand{\ba}{\begin{array}}
\newcommand{\ea}{\end{array}}

\newcommand{\bea}{\begin{eqnarray}}
\newcommand{\eea}{\end{eqnarray}}

\newcommand{\su}{\mbox{su}}
\newcommand{\so}{\mbox{so}}
\newcommand{\SO}{\mbox{SO}}

\def\E{{\cal E}}
\def\L{{\cal L}}

\def\R{{\cal R}}
\def\A{{\cal A}}
\def\B{{\cal B}}

\def\hpartial{\hat{\partial}}
\def\g{g_{\scriptscriptstyle
{Y\!M}}}

\def\tr{{\rm tr}}
\def\Tr{{\rm Tr}}
\def\trN{{\rm tr}_{\!{\scriptscriptstyle N}}}

\def\I_M{{I_{\scriptscriptstyle M\times M}}}

\def\N{{\cal  N}}

\title{3D $\N=2$ massive  super Yang-Mills and membranes/D2-branes in a curved background
}

\author{Seungjoon Hyun,${}^{\ast}$ Jeong-Hyuck Park$^{\dagger}$ and Sang-Heon
Yi$^{\ast}$\\
${}^{\ast}$Institute of Physics and Applied Physics, Yonsei University, Seoul 120-749, Korea\\
${}^{\dagger}$Korea Institute for Advanced Study,   Seoul 130-012, Korea\\
\\
E-mail~:~\email{hyun@phya.yonsei.ac.kr}, \email{jhp@kias.re.kr}, \email{shyi@phya.yonsei.ac.kr}}

\abstract{We present a three dimensional novel massive $\N=2$ super
  Yang-Mills action as  a low energy effective worldvolume description
  of  the D2-branes on a pp-wave. The action contains the Myers term,
  mass  terms for three Higgs, and terms mixing the electric fields
  with  other two Higgs.
 We derive the action in three different  ways, from the M-theory
  matrix model, from  the supermembrane action, and from the
  Dirac-Born-Infeld action. We verify the consistent mutual  agreement
  and comment how each approach is  complementary to another.
  In particular,  we   give the eleven dimensional geometric interpretation of the vacua in
  the worldvolume theory as  the membranes tilted to the eleventh direction  with the giant
gravitons around.}


\preprint{KIAS-P03001\\
hep-th/0301090}

\begin{document}

\section{Introduction and summary}
D-branes are a cornerstone to show that the five perturbative superstring theories in ten dimensions belong
to  the    unique eleven dimensional theory or the   M-theory \cite{D-brane}. Although  the worldvolume
action for the D-brane is generically given by the Dirac-Born-Infeld action, the precise form of  its
supersymmetric non-Abelian generalization has not been yet known, especially in the general curved
background. One can merely expect that,  in the generic background, the leading term of such,  if any,
generalized DBI action  will correspond to a certain modification of the super Yang-Mills, since in the flat
background it should be the
ordinary super Yang-Mills.\\

Recently, there have been much interests in the string/M-theory in the maximally supersymmetric ten/eleven dimensional  pp-wave backgrounds. Strings
on the 10D pp-wave are exactly solvable
\cite{Berenstein:2002jq,Metsaev:2001bj,Metsaev:2002re,Brecher:2002ar,Alishahiha:2002nf,Hyun:2002wu,Hyun:2002wp}, and the exact form of the M-theory
matrix model in the 11D pp-wave background is available now, thanks to Berenstein, Maldacena and Nastase (BMN)
\cite{Berenstein:2002jq} (see also \cite{Kim:2002cr,Lee:2002vx}).\\

One characteristic feature of the string theory in the pp-wave
background is that the string modes are all massive,
\begin{equation}
E_{n}=\sqrt{\mu^{2}+n^{2}/(\alpha^{\prime}p_{+})^{2}}\,,
\end{equation}
where $\mu$ is the characteristic mass parameter in the pp-wave geometry. Consequently,  the worldvolume
descriptions of the D-branes in the low energy limit are expected to be given by `massive' gauge theories. It
is, thus, important to understand how to realize the theory of massive vector supermultiplets while
maintaining the gauge invariance  \cite{HyunPark,Metsaev:2002sg,Metsaev:2003bf,Bonelli:2002mb}.  The
  main motivation of
the present paper is to construct such a massive supersymmetric
gauge theory as a low energy worldvolume description of the membranes
or D2-branes in the pp-wave background.\\

The BMN matrix model corresponds to a mass deformation of the BFSS matrix model
\cite{Banks:1996vh,Susskind:1997cw,Sen,Seiberg}, still maintaining the maximal thirty two supersymmetries.
Due to the  existing   mass parameter, $\mu$, the BMN  matrix model presents many distinctive features, not
shared by the BFSS matrix model. Among others, the supersymmetry transformations  have the explicit time
dependency. Accordingly the supercharges do not commute with the Hamiltonian, and the corresponding
supersymmetry algebra is identified as the special unitary Lie superalgebra, $\su(2|4;2,0)$ for $\mu>0$ or
$\su(2|4;2,4)$ for $\mu<0$, of which the complexification is $\mbox{A}(1|3)$.  Refs.~\cite{KimPark,Keshav}
contain the complete classification of its representations, including the quantum BPS multiplets as the
`atypical' representations.  The classical counterparts of the quantum BPS states are the bosonic
configurations which are the solutions of the BPS equations. In \cite{Park:2002cb}, all the BPS equations
were obtained which correspond to the  quantum BPS states preserving the various fractions of the dynamical
supersymmetry, $2/16$, $4/16$, $8/16$, $16/16$. For the discussion of the perturbative aspects of the BPS
states, see
\cite{Dasgupta:2002hx,Kim:2002if,Maldacena:2002rb}.\\

One characteristic  feature of the generic BPS configurations is that, either they are rotating  with a constant frequency, or static but curved
\cite{Park:2002cb,Hyun:2002cm,Bak:2002rq,Mikhailov:2002wx}. In any case, it is an artifact of the coordinate choice that the branes, especially of
the infinite size, are rotating. In fact, adopting a  comoving rotating coordinate system, one can reformulate the matrix model such that the BPS
configurations are static. Expanding the matrix model around the static BPS configuration leads to a non-commutative gauge theory, and taking the
commutative limit one can obtain the low energy effective worldvolume action for the branes. In this way, the worldvolume action for the
longitudinal five branes or the D4-branes in the pp-wave background was obtained in our previous work \cite{HyunPark}. The resulting action is a five
dimensional massive $\N=1/2$ super Yang-Mills coupled to the K\"{a}hler-Chern-Simons term. In particular, the gauge fields acquire mass through the
K\"{a}hler-Chern-Simons term \cite{NairKCS}. \\

Another interesting BPS solution found in \cite{Hyun:2002cm}, which is the main theme of the present paper,  is the rotating flat membranes
preserving four supersymmetries. In contrast to  the longitudinal five branes or other known BPS solutions, this configuration preserves certain
nontrivial four combinations of the dynamical and kinematical supersymmetries. Since the kinematical supercharges and the dynamical supercharges in
the BMN matrix model have different quantum numbers for the Hamiltonian,  such configurations do not correspond to the energy
eigenstates which have been classified in \cite{Park:2002cb}.\\

In the present paper,  we study  the above membrane configuration in three different  ways,  from the
M-theory matrix model, from the supermembrane action, and from the Dirac-Born-Infeld action.  We confirm the
existence of the supersymmetric membrane configuration and derive its low energy effective worldvolume
action in each setup. We verify the consistent mutual  agreement and   comment how each approach is
complementary to another. In particular,  after constructing the precise dual relation between the  field
strength  and a compact scalar, we  give the eleven
dimensional geometric interpretation of  the vacua in the worldvolume theory  as  a membrane tilted to the eleventh direction.\\

Our resulting worldvolume action is a three dimensional massive $\N=2$ super Yang-Mills which contains the
Myers term, mass terms for three Higgs, and terms mixing the electric fields with other two Higgs. Notably
the last ones make the gauge fields massive, which is quite different from  the well-known mechanism through
the Chern-Simons term \cite{Siegel:1979fr}. We write the action here, as a power series of the mass
parameter, $\mu$,
\begin{equation}
{\cal S}=\displaystyle{\frac{1}{\g^{2}} \int\,dx^{3}}~
\L_{0}+\mu\L_{1}+\mu^{2}\L_{2}\,, \label{M2action}
\end{equation}

\begin{equation}
\ba{l}{\L_{0}=\tr_{\!{\scriptscriptstyle N}}\Big[-\quarter
  F_{\mu\nu}F^{\mu\nu}-\half  D_{\mu}\phi_{a}D^{\mu}\phi_{a}+\quarter
[\phi_{a},\phi_{b}]^{2}- i\half\psi^{\dagger}\gamma^{\mu} D_{\mu}
 \psi- \half\psi^{\dagger}\gamma^{a}[\phi_{a},\psi]\Big]\,,}\\
{}\\
\L_{1}=\textstyle{\frac{1}{\sqrt{2}}}\,\tr_{\!{\scriptscriptstyle N}}
\Big[\textstyle{\frac{1}{3}} (\phi_{4}F_{01}+ \phi_{3}F_{02})
+\textstyle{\frac{1}{6}}\epsilon^{pq}\phi_{p}D_{0}\phi_{q}
  -i\textstyle{\frac{1}{3}}\epsilon^{rst}\phi_{r}\phi_{s}\phi_{t}+
i\textstyle{\frac{1}{24}}\psi^{\dagger}\Pi\psi \Big]\,,\\
{}\\
{\L_{2}=-\,\half(\textstyle{\frac{1}{3\sqrt{2}}})^{2}\,\tr_{\!{\scriptscriptstyle
  N}}\left(\phi_{7}^{\,2}+\phi_{8}^{\,2}+\phi_{9}^{\,2}\right)\,.}
\ea \label{M2LLL}
\end{equation}
where $\mu,\nu=0,1,2$, $a=3,4,5,6,7,8,9$, $p=5,6$, $r=7,8,9$,
$\epsilon^{56}=\epsilon^{789}=1$,  and
\begin{equation}
\Pi=(\gamma^{14}+\gamma^{23}-\gamma^{56}+3\gamma^{789})\,.
\end{equation}
\newpage

The organization of the present paper and the summary of the  results are as follows. In section
\ref{setup}, we give the basic setup for both the matrix model and the supergravity, mainly to establish the
notations and the conventions. In particular, introducing the rotating coordinate system, we
reformulate the BMN matrix model.\\

In section \ref{object}, we identify the  static membrane configurations preserving four supersymmetries. In the reformulated matrix model, we
explicitly  construct the solution and find that the preserved supersymmetries are linear combinations of the kinematical and dynamical
supersymmetries.  In the supergravity setup, we perform the probe analysis and show that  a  membrane spanning the  $(x^{0},x^{1},x^{2})$ directions
is supersymmetric (\textit{cf.} \cite{Skenderis:2002vf,Kim:2002tj}).\\

In section \ref{derivation}, we derive the low energy effective worldvolume actions for the M2 and D2 branes
in three different ways. From the matrix model, we first get the non-commutative version of  the non-Abelian
action, (\ref{M2action}), (\ref{M2LLL}), and then  take the commutative limit. From the supermembrane action,
in the  low energy limit,  we obtain a supersymmetric scalar  action, (\ref{SM2}), while from the
Dirac-Born-Infeld action we acquire a bosonic  massive gauge theory action, (\ref{SD2}). The comparison among
the results is done in the subsection \ref{compare}.  We verify the consistent mutual agreement.  The latter
two actions are shown to be equivalent by constructing the dual relation between the  field strength
and a compact scalar, (\ref{dual}).\\

In section \ref{wsusy}, we identify the worldvolume supersymmetries from the matrix model and from the
supergravity,  respectively. In the matrix model, we first observe that in the commutative limit,
transformations of some dynamical supersymmetries become singular. By imposing two constraints on the
sixteen component dynamical supersymmetry parameter, the singularity is removed and only four
supersymmetries survive. In the supergravity setup, the same worldvolume supersymmetries are identified as
the combinations of the spacetime supersymmetry and the $\kappa$-symmetry which  preserve the
$\kappa$-symmetry fixing as well as the static gauge choice of the worldvolume coordinates. The subsection
\ref{sas} presents the relevant 3D $\N=2$ supersymmetry algebra, (\ref{susyalge}), and discusses the
existing three  supermultiplets in the massive super Yang-Mills which are characterized by the different
energy spectra. In the last subsection \ref{ssBPS}, we write the BPS equations of the worldvolume theory
which describe the bosonic configurations preserving  all the four supersymmetries. Due to the novel
structure of the supersymmetry algebra, these BPS equations are not trivial. In particular, the solutions of
the vacua are given by the constant fuzzy spheres formed by the last three Higgs,
$(\phi^{7},\phi^{8},\phi^{9})$ and arbitrary two vevs of $(\phi^{3},\phi^{4})$. Utilizing the dual relation
between the field strength and the compact scalar, we give the eleven dimensional geometric interpretation
of the vacua. Namely they correspond to the membranes tilted to the eleventh direction with the giant
gravitons around.\\

The appendix  contains some useful formulae and explicit forms of the supercharge, $R$-symmetry charges and  central charges in the worldvolume
theory.

\newpage
%%%%%%%%%%%%%%%%%%%%%%%%%%%%%%%%%%%%%%%%%%%%%%%%%%%%%%%%%%%%%%%%%%%%%%%%%%%%%%%%%%%%%%%%%%%%%%%%%%%%%%%%%%%%%%
%%%%%%%%%%%%%%%%%%%%%%%%%%%%%%%%%%%%%%%%%%%%%%%%%%%%%%%%%%%%%%%%%%%%%%%%%%%%%%%%%%%%%%%%%%%%%%%%%%%%%%%%%%%%%%

\section{Setup\label{setup}}
In this section, we give the basic setup for both the matrix model and the supergravity.  First, we reformulate  the BMN matrix model  in a rotating
coordinate system. In the second part, we write the basic formalism for the supermembrane action on the pp-wave.
\subsection{Matrix model in a rotating  coordinate system}
In \cite{Hyun:2002cm}, it was shown that the  BMN matrix model admits a rotating membrane  preserving four supersymmetries,  each of which is a
linear combination of the dynamical  and kinematical supersymmetries.  It is an artifact of the coordinate choice that the membranes rotate with a
constant frequency. The original BMN matrix model  was written    in a maximally symmetric coordinate system, where the pp-wave metric is of the
form \cite{Kowalski-Glikman,Figueroa-O'Farrill1,FO2},
\begin{equation}
\ba{l}ds^{2}=-2dx^{+}dx^{-}-\Big[(\textstyle{\frac{\mu}{6}})^{2}(x_{1}^{2} +\cdots+x_{6}^{2})+
(\textstyle{\frac{\mu}{3}})^{2}(x_{7}^{2}+x_{8}^{2}+x_{9}^{2})\Big]dx^{+}dx^{+}
+\displaystyle{\sum_{A=1}^{9}}\,dx^{A}dx^{A}\,,\\
{}\\
F_{+789}=\mu\,, \ea\label{pp-wave}
\end{equation}
with the isometry group, $\SO(6)\times\SO(3)$. Reformulating the matrix model in a  less symmetric  but `comoving' coordinate system, one can obtain
the `static' membrane configuration.  Explicitly we replace the first six coordinates, $x_{1},x_{2},x_{3},x_{4},x_{5},x_{6}$, by the
$\SO(2)\times\SO(2)\times\SO(2)$ rotating ones,
\begin{equation}
\ba{cc} x_{1}\rightarrow \cos(\mu x^{+}/6)x_{1}+\sin(\mu
x^{+}/6)x_{4}\,,~~~&~~~x_{4}\rightarrow
\cos(\mu x^{+}/6)x_{4}-\sin(\mu x^{+}/6)x_{1}\,,\\
{}&{}\\
x_{2}\rightarrow \cos(\mu x^{+}/6)x_{2}+\sin(\mu
x^{+}/6)x_{3}\,,~~~&~~~ x_{3}\rightarrow \cos(\mu
x^{+}/6)x_{3}-\sin(\mu x^{+}/6)x_{2}\,,\\
{}&{}\\
x_{5}\rightarrow \cos(\mu x^{+}/6)x_{5}-\sin(\mu
x^{+}/6)x_{6}\,,~~~&~~~ x_{6}\rightarrow \cos(\mu
x^{+}/6)x_{6}+\sin(\mu x^{+}/6)x_{5}\,, \ea\label{rota}
\end{equation}
so that the metric of the eleven dimensional  pp-wave background, (\ref{pp-wave}), is,  in the new coordinate
system, of the form
\begin{equation}
\ba{ll}ds^{2}=&-2dx^{+}dx^{-}-\frac{\mu}{3}(x_{1}dx_{4}-x_{4}dx_{1}+x_{2}dx_{3}-x_{3}dx_{2}-x_{5}dx_{6}+x_{6}dx_{5})dx^{+}\\
{}&{}\\
{}& -\left(\frac{\mu}{3}\right)^{2}(x_{7}^{2}+x_{8}^{2}+x_{9}^{2}) dx^{+}dx^{+}+\displaystyle{\sum_{A=1}^{9}}\,dx^{A}dx^{A}\,. \ea\label{pp-wave2}
\end{equation}
The rotation of the $(5,6)$ plane is not necessary to obtain the static configuration. However,  it makes
the supercharges commute with the Hamiltonian in the worldvolume theory, as is the case for the worldvolume
theory  of the
longitudinal five branes \cite{HyunPark}.\\

The corresponding M-theory matrix model on this background is obtained from the original  BMN matrix model by incorporating  the above time dependent
rotations. With $t\equiv x^{+}$,  the transformations of the bosons are essentially the same as above,
\begin{equation}
\ba{ll} X_{1}\rightarrow \cos(\mu t/6)X_{1}+\sin(\mu
t/6)X_{4}\,,~~&~~\mbox{etc.}\ea\label{XROTATION}
\end{equation}
while those of the fermions read, from the standard Lorentz transformation rule,
\begin{equation}
\psi~\rightarrow~\displaystyle{e^{\frac{\mu}{12}(\gamma^{14}+\gamma^{23}-\gamma^{56})t}\psi\,.}\label{PSIROTATION}
\end{equation}

The  modified, but nevertheless equivalent,  M-theory matrix model on the  fully supersymmetric  pp-wave background spells\footnote{For  the
derivation of the original  BMN matrix model either from the supergraviton action or from the Polyakov type supermembrane action, we refer
\cite{Berenstein:2002jq} and \cite{Dasgupta:2002hx} respectively.}
 with a mass parameter,
$\mu$,
\begin{equation}
{\cal
S}=\displaystyle{\frac{l_{p}^{6}}{R^{3}}}\displaystyle{\int\,dt}~
\L_{0}+\mu\L_{1}+\mu^{2}\L_{2}\,, \label{Maction}
\end{equation}

\begin{equation}
\ba{l} \L_{0}=\Tr\!\left(\half D_{t}X^{A}D_{t}X_{A}+\quarter
[X^{A},X^{B}]^{2}+i\half
\psi^{\dagger}D_{t}\psi-\half\psi^{\dagger}\gamma^{A}[X_{A},\psi]\right)\,,\\
{}\\
\L_{1}=\Tr\!\left[-\textstyle{\frac{1}{6}}J^{ab}X_{a}D_{t}X_{b} -i\textstyle{\frac{1}{3}}\epsilon^{rst}X_{r}X_{s}X_{t}
+i\textstyle{\frac{1}{24}}\psi^{\dagger}(\gamma^{14}+\gamma^{23}-\gamma^{56}+3\gamma^{789})\psi \right]\,,\\
{}\\
\L_{2}=-\,\half\,(\textstyle{\frac{1}{3}})^{2}\Tr\!\left(X_{7}^{\,2}+X_{8}^{\,2}+X_{9}^{\,2}\right)\,, \ea\label{MLagrangian}
\end{equation}
where   $a,b=1,2,3,4,5,6$, $~r,s,t=7,8,9$, $~A,B=1,2,\cdots,9$ and $J^{ab}$ is a skew-symmetric $6\times 6$ constant two form of which the
non-vanishing components are $J^{14}=J^{23}=J^{65}=1$ only, up to the anti-symmetric property. In the present paper, we adopt generic Euclidean nine
dimensional gamma matrices, $\gamma^{A}=(\gamma^{A})^{\dagger}$, $\gamma^{12\cdots 9}=1$. Namely we do not adopt the usual real and symmetric
Majorana representation. Accordingly there exits a nontrivial $16\times 16$ charge conjugation matrix, ${C}$,
\begin{equation}
\begin{array}{ll}
(\gamma^{A}){}^{T}=(\gamma^{A}){}^{\ast}={C}^{-1}\gamma^{A}{C}\,,~~~&~~~C=C^{T}=(C^{\dagger})^{-1}\,.
\end{array}\label{chargeC}
\end{equation}
The spinors, $\psi$,  satisfy the Majorana condition leaving eight
independent  complex components,
\begin{equation}
\psi={C}\psi^{\ast}\,. \label{Majoranacondition}
\end{equation}
The covariant derivatives are in our convention, $D_{t}{\cal O}=\frac{d~}{dt}{\cal O}-i[A_{0},{\cal O}]$ so that $X$ and $A_{0}$ are of the mass
dimension one, while $\psi$ has the mass dimension ${3}/{2}$. Compared to the original BMN matrix model, the quadratic mass terms for the  first six
bosonic coordinates are absent.
\newline

The dynamical or linearly realized supersymmetry transformations
are
\begin{equation}
\ba{l} \delta A_{0}=i\psi^{\dagger}\E(t)\,,~~~~~~~~~~~\delta X^{A}=i\psi^{\dagger}\gamma^{A}\E(t)\,,\\
{}\\
\delta\psi=\Big[D_{t}X^{A}\gamma_{A}-i\half
[X^{A},X^{B}]\gamma_{AB}-
\textstyle{\frac{\mu}{3}}(X^{7}\gamma_{7}+X^{8}\gamma_{8}+X^{9}\gamma_{9})\gamma^{789}\\
{}\\
~~~~~~~~~~~+\textstyle{\frac{\mu}{6}}(X^{1}\gamma_{1}+X^{4}\gamma_{4})(\gamma^{789}-\gamma^{14})+
\textstyle{\frac{\mu}{6}}(X^{2}\gamma_{2}+X^{3}\gamma_{3})(\gamma^{789}-\gamma^{23})\\
{}\\
~~~~~~~~~~~+\textstyle{\frac{\mu}{6}}(X^{5}\gamma_{5}+X^{6}\gamma_{6})(\gamma^{789}+\gamma^{56})
\Big]\E(t)\,, \label{susytr}\ea
\end{equation}
where
\begin{equation}
\ba{cc}
\displaystyle{\E(t)=e^{\frac{\mu}{12}(-\gamma^{14}-\gamma^{23}+\gamma^{56}+\gamma^{789})t}\E\,,}~~~&~~~
\E={C}\E^{\ast}\,,\ea \label{Et}
\end{equation}
and $\E$ is an arbitrary  sixteen component constant spinor.\\

In addition,  there is  the kinematical or the non-linearly realized supersymmetry,
\begin{equation}
\ba{ccc} \delta A_{0}=\delta X^{A}=0\,, ~~~~&~~~~
\displaystyle{\delta\psi=e^{-\frac{\mu}{12}(\gamma^{14}+\gamma^{23}-\gamma^{56}+3\gamma^{789})t}\E^{\prime}\,,}
~~~&~~~\E^{\prime}={C}\E^{\prime}{}^{\ast}\,.\ea\label{Eprimet}
\end{equation}

\subsection{Supermembrane action on the pp-wave}
Here we briefly review the formalism of the supermembrane action given in \cite{deWitpp}. Mostly following the conventions therein, except
$\epsilon^{012}=1$, we denote the curved  space indices by $\tilde{M} = (M, \alpha )$, and the tangent  space indices by   $ \tilde{A} =
(\hat{R},\hat{a})$, while  $\mu, \nu=0,1,2$ are the worldvolume indices of the supermembrane.\\

Using the superspace embedding  coordinates\footnote{$\theta$ is a 32-component  spinor satisfying the Majorana condition,
 (\ref{11DMajorana}).},  $Z^{\tilde{M}}(\xi) = (x^M(\xi), \theta^{\alpha}(\xi))$, the supermembrane action is given by
\be S_{M2} = - T_{M2}\int d\xi^{3} \sqrt{-h(Z(\xi))}  + T_{M2}\int B \,. \label{M2}\ee
Here $h$ is the determinant of the induced worldvolume metric,
\be  h=\det h_{\mu\nu}\,,\qquad h_{\mu\nu}=\Pi^{~\hat{R}}_\mu\Pi^{~\hat{S}}_{\nu} \eta_{\hat{R}\hat{S}} \,, \ee written in terms of the pull-back,
$\Pi^{~\tilde{A}}_{\mu}$,  of the supervielbein, $E^{\tilde{A}}_{\tilde{M}}$,
\be \Pi^{~\tilde{A}}_\mu =
\partial_\mu x^M E^{\tilde{A}}_M +
\partial_\mu\theta^\alpha E^{\tilde{A}}_\alpha\,.\ee
The three-form superfield, $B$,  gives   the Wess-Zumino term,
\be
 T_{M2}\int B = T_{M2}\int d\xi^3
\frac{\epsilon^{\mu\nu\rho}}{6\sqrt{-h}} \Pi^{~\tilde{A}}_\mu\Pi^{~\tilde{B}}_{\nu}\Pi^{~\tilde{C}}_{\rho}B_{\tilde{C}\tilde{B}\tilde{A}} \,, \ee
and  can be expanded  in terms of $\theta$  and $\bar{\theta} = i\theta^\dagger\Gamma^0$ \cite{deWit:1998yu},
\begin{equation}
B = -\frac{1}{6} e^{\hat{R}}\wedge e^{\hat{S}}\wedge e^{\hat{U}}C_{\hat{R}\hat{S}\hat{U}} + \frac{1}{2}e^{\hat{R}}\wedge e^{\hat{S}}\wedge
\bar{\theta} \Gamma_{\hat{R}\hat{S}}D\theta + {\cal O}(\theta^4)\,, \label{Bform}
\end{equation}
where
\be\ba{ll}D\theta = d\theta -\frac{1}{4}\omega^{\hat{R}\hat{S}}\Gamma_{\hat{R}\hat{S}}\theta + e^{\hat{R}}T_{\hat{R}}^{NPQR} F_{NPQR}\theta
\,,~~~&~~~ T_M^{NPQR} = \frac{1}{288}(\Gamma_M^{\; NPQR}
 -8\delta^{\; [N}_M\Gamma^{PQR]})\,.
\label{Dtheta} \ea\ee
The $\kappa$-symmetry of the supermembrane action is given by
\be \delta_{\kappa} Z^{\tilde{M}} E_{\tilde{M}}^{\hat{R}} = 0\, ,
 \qquad \delta_{\kappa}Z^{\tilde{M}}
 E_{\tilde{M}}^{\hat{a}} = (1+\Gamma)^{\hat{a}}{}_{\hat{b}}{}\kappa^{\hat{b}} \,, \label{SetKappa}\ee
where $\kappa(\xi)$ is an arbitrary  local fermionic parameter and $\Gamma$ is the projection matrix, \be
  \Gamma =
\frac{1}{6}\frac{\epsilon^{\mu\nu\rho}}{\sqrt{-h}}
 \Pi^{~\hat{R}}_\mu\Pi^{~\hat{S}}_{\nu}\Pi^{~\hat{U}}_{\rho}
\Gamma_{\hat{R}\hat{S}\hat{U}}\,,
 \qquad \epsilon^{012}=1\,,\label{KG}
\ee
satisfying
\be
\tr \Gamma = 0\,, \qquad \Gamma^2=1\,.
\ee

The component form of the supermembrane action in the  general background is known only up to $\theta^2$ order~\cite{deWitpp}. However, the explicit
forms  in the  maximally supersymmetric $AdS_4\times S^7$ and $AdS_7\times S^4$ have been determined to all orders,  using the coset
method~\cite{deWit:1998yu}. The corresponding supervielbein for these spaces is
\bea E & =& D\theta + \sum_{n=1}^{16}\frac{1}{(2n+1)!}{\cal
M}^{2n}D\theta \, , \label{E1}\\
\nonumber\\
 E^{\hat{R}} & = & e^{\hat{R}} +
\bar{\theta}\Gamma^{\hat{R}}D\theta +
2\sum_{n=1}^{15}\frac{1}{(2n+2)!}\bar{\theta}\Gamma^{\hat{R}}{\cal
M}^{2n}D\theta \, ,\label{E2}\eea
where
\bea
 ({\cal M}^2)^{\hat{a}}{}_{\hat{b}} &=& 2(T_M^{NPQR}\theta)^{\hat{a}}
  F_{NPQR} (\bar{\theta}\Gamma^M)_{\hat{b}} \nonumber \\
  &&\nonumber\\
  &&-\frac{1}{288}(\Gamma_{MN}\theta)^{\hat{a}}
  \Big[\bar{\theta}(\Gamma^{MNPQRS}
  F_{PQRS}+24F^{MNPQ}\Gamma_{PQ})\Big]_{\hat{b}}\,.
  \eea
Since the maximally supersymmetric pp-wave can be obtained by taking a Penrose limit of the maximally
supersymmetric $AdS\times S$ spaces, the result above is still valid for  the pp-wave geometry
\cite{Dasgupta:2002hx}.\\

Rotating  $x^{1},x^{2},x^{3},x^{4}$ coordinates as in (\ref{rota}) and transforming the $x^{-}$ coordinate as\footnote{Note that the shift of $x^{-}$
coordinate would result in adding a total derivative term in the M-theory matrix model.}  $x^{-}\rightarrow x^{-}-(\mu/6)(x^1 x^4 + x^2 x^3)$,  we
rewrite the pp-wave geometry  (\ref{pp-wave}), with $x^{\pm}=\frac{1}{\sqrt{2}}(t\pm y)$,
\bea ds^2_{11} &=& -(1+H/2)\Big[dt+\frac{H/2}{1+H/2} dy\Big]^2 + \frac{1}{1+H/2}dy^2 +
\Big[dx^1+\frac{\mu}{3\sqrt{2}}x^4(dt+dy)\Big]^2  \nonumber \\
&&\mbox{}+\Big[dx^2+\frac{\mu}{3\sqrt{2}}x^3(dt+dy)\Big]^2  + \sum_{l=3}^9dx^ldx^l \,,
\nonumber \\ \nonumber
\\
F_{t789}&=&F_{y789}=\frac{\mu}{\sqrt{2}}\,, \label{metric} \eea
where \be H=
\textstyle{(\frac{\mu}{6})^{2}(x_5^2+x_6^2)+(\frac{\mu}{3})^{2}(x_3^2+x_4^2+x_7^2+x_8^2+x_9^2)\,.}
\ee

Appendix \ref{usefulA} contains the explicit forms of the  bosonic vielbein and spin connections as well as our choice of the 11D gamma matrix
representation which utilizes  the 9D gamma matrix used in the M-theory matrix model setup.

%%%%%%%%%%%%%%%%%%%%%%%%%%%%%%%%%%%%%%%%%%%%%%%%%%%%%%%%%%%%%
%%%%%%%%%%%%%%%%%%%%%%%%%%%%%%%%%%%%%%%%%%%%%%%%%%%%%%%%%%%%%

\section{BPS membranes preserving four supersymmetries\label{object}}
In this section, we discuss the existence of the BPS membrane configurations which preserve four
supersymmetries, in each framework. First, in the matrix model setup, we obtain the static BPS membrane
solution, and show that only four supersymmetries are unbroken. They  are given by the linear combination of
the dynamical and kinematical supersymmetries. Then, in the supergravity setup, we perform the relevant
probe analysis to identify the corresponding membrane configuration and the four supersymmetries.

\subsection{Matrix model analysis}
We consider the following static flat membrane configurations,
\begin{equation}
\ba{ccc}X_{1}=i\hat{\partial}_{1}\,,~~~&~~~X_{2}=i\hat{\partial}_{2}\,,
~~~&~~~A_{0}={X_{A}}=0\,,~~~A=3,4,\cdots,9\,.\ea\label{M2sol}
\end{equation}
Here the operators, $\hat{\partial}_{i}$'s are related to the coordinates of a  non-commutative plane,
\begin{equation}
\ba{cc}x^{1}=i\theta\hpartial_{2}\,,~~~~&~~~~x^{2}=-i\theta\hpartial_{1}\,,\ea\label{M2sol2}
\end{equation}
such that
\begin{equation}
\ba{ccc} {}[X_{1},X_{2}]=i\frac{1}{\theta}\,,~~~&~~~
[x^{1},x^{2}]=i\theta\,,~~~&~~~
[\hpartial_{i},x^{j}]=\delta_{i}{}^{j}\,.\ea\label{non-com}
\end{equation}
This relation  gives a set of harmonic oscillators, and   the most general irreducible representation is  specified by the superselection rule on
the number of the ground states which we denote by $N$.  Thus, the Hilbert space, ${\cal H}$, on which the infinite matrices act decomposes as a
direct product of a harmonic oscillator Hilbert space, $H_{h.o.}$ and an $N$ dimensional vector space, $V_{N}$,
\begin{equation}
{\cal H}=H_{h.o.}\otimes  V_{N}\,.
\end{equation}
Explicitly, using the bra and ket notations, one can  regroup the states in the Hilbert space  \cite{JHPcomments},
\begin{equation}
\ba{cccc} |n,s\rangle\,,~~&~~n=0,1,\cdots,\infty\,,~&~s=1,2,\cdots,N\,, \ea
\end{equation}
so that the  creation and annihilation operators are respectively,
\begin{equation}
\ba{ll} \displaystyle{\sum_{n,s}\sqrt{n+1}|n+1,s\rangle\langle
n,s|\,,}~~~&~~~
\displaystyle{\sum_{n,s}\sqrt{n+1}|n,s\rangle\langle n+1,s|\,.}
\ea
\end{equation}
Of course, this represents $N$ parallel membranes which, we show,  preserve four supersymmetries.\\

To see that the  configuration preserves four supersymmetries, we
pay attention to the supersymmetry transformation of the fermions
which, in the present  case, reduces to
\begin{equation}
\delta\psi=\frac{1}{\theta}\gamma^{12}e^{\frac{\mu}{12}(-\gamma^{14}-\gamma^{23}+\gamma^{56}+\gamma^{789})t}\E
+e^{-\frac{\mu}{12}(\gamma^{14}+\gamma^{23}-\gamma^{56}+3\gamma^{789})t}\E^{\prime}\,,
\end{equation}
where the first and second parts are dynamical and kinematical supersymmetry transformations respectively. Requiring it to vanish,  we obtain
\begin{equation}
\E^{\prime}=-\frac{1}{\theta}\gamma_{12}e^{\frac{\mu}{6}(-\gamma^{14}-\gamma^{23}+2\gamma^{789})t}\E\,.
\end{equation}
Since the left hand side is time independent, the Killing spinor, $\E$, must satisfy
\begin{equation}
(\gamma^{14}+\gamma^{23}-2\gamma^{789})\E=0\,.\label{GGGE0}
\end{equation}
By multiplying  $\gamma^{1234},\gamma^{1456},\gamma^{2356}$ to the left and using $\gamma^{12\cdots 9}=1$, one can show that the constraint  is
actually  equivalent to
\begin{equation}
\gamma^{14}\E=\gamma^{23}\E=\gamma^{56}\E=\gamma^{789}\E\,.\label{GGE}
\end{equation}
In a more concise form, this is again equivalent to
\begin{equation}
\ba{ll}\Omega\E=\E\,,~~~&~~~~\Omega=\quarter(1-\gamma^{1234}-\gamma^{1456}-\gamma^{2356})\,.\ea\label{Omega}
\end{equation}
$\Omega$ is a   projection matrix for the Killing spinors  \cite{Park:2002cb,jhpBPS} satisfying,
\begin{equation}
\ba{cccc} \Omega^{\dagger}=\Omega\,,~~~&~~~ C\Omega^{\ast}C^{-1}=\Omega\,,~~~&~~~ \Omega^{2}=\Omega\,,~~~&~~~\tr\Omega=4\,,\ea
\end{equation}
and, to agree with (\ref{GGE}),
\begin{equation}
\gamma^{14}\Omega=\gamma^{23}\Omega=\gamma^{56}\Omega=\gamma^{789}\Omega\,.\label{GGO}
\end{equation}
Thus, the configuration preserves four supersymmetries, each of
which is  a linear combination of the dynamical and kinematical
supersymmetries given by
\begin{equation}
\E^{\prime}=-\frac{1}{\theta}\gamma_{12}\E\,.\label{KD}
\end{equation}

%%%
%% In general, the Killing spinors in the supersymmetry transformations form a kernel space.
%%Analyzing  the projection matrix to the kernel,   one can
%%obtain in a systematic way all the possible sets of the BPS equations of various
%%unbroken supersymmetry fractions \cite{Park:2002cb,jhpBPS}.
%%%

In the ordinary BFSS matrix model  or the $\mu=0$ flat
 background case, the same membrane configuration, (\ref{M2sol}) and (\ref{non-com}),
preserves sixteen supersymmetries out of thirty two. They  are given by  the linear combinations of the kinematical and dynamical supersymmetries,
(\ref{KD}),  without any constraint on $\E$.\\

Finally, from (\ref{susytr}), it is worth to note that the membrane configuration can be shifted to the third and fourth directions still preserving
the four supersymmetries,
\begin{equation}
\ba{ll}X^{3}=c^3\,1\,,~~~~&~~~~X^{4}=c^{4}\,1\,.\ea\label{shift}
\end{equation}
In the subsection \ref{ssBPS}, we will see that this configuration, in fact, corresponds to membranes tilted to the 11th direction.

\subsection{Probe analysis}
%%%%%%%%%%%%%%%%%%%%%%%%%%%%%%%%%%%%%%%
Here we count the number of supersymmetries  a M2-brane probe preserves. In the probe analysis, we only consider the rigid flat M2-brane. We will
take into account its fluctuations when we consider
the worldvolume supersymmetry in the subsection \ref{SSS}.\\

The supersymmetry variation of the gravitino, $\psi_M$, is
\be \delta_\eta \psi_M = (\partial_M -
\frac{1}{4}\omega^{\hat{R}\hat{S}}_M \Gamma_{\hat{R}\hat{S}}+
T_M^{NPQR}F_{NPQR} )\eta \,. \ee
The Killing spinor, $\eta$,  satisfying $\delta_\eta \psi =0$,  for the given pp-wave geometry,
(\ref{metric}),  is of the form,
\be \eta = (\cosh\ln(1+H/2)^{1/4}+\Gamma^{\hat{+}\hat{-}}
\sinh\ln(1+H/2)^{1/4}) (1-\sum_{n=1}^{9}x^n\Omega_n)e^{-x^+\Omega_+}
\eta_0\,,   \label{Killing}\ee
%
%where $x^\pm \equiv \frac{1}{\sqrt{2}}(t\pm z).$
where  $\Gamma^{\hat{\pm}}=\frac{1}{\sqrt{2}}(\Gamma^{\hat{t}}\pm \Gamma^{\hat{y}})$,  $~~\eta_0$ is an
arbitrary  32-component constant spinor, and \bea &\Omega_+& =
\textstyle{\frac{\mu}{12}}(\Gamma^{\hat{+}\hat{-}\hat{7}\hat{8}\hat{9}}
         +\Gamma^{\hat{1}\hat{4}}
  +\Gamma^{\hat{2}\hat{3}}-2\Gamma^{\hat{7}\hat{8}\hat{9}})\,,\label{O+}
                \\
         \nonumber       \\
 &\Omega_n& = \left\{
\begin{array}{ll}
                  \frac{\mu}{12}\Gamma^{\hat{+}\hat{4}}
                  (1-\Gamma^{\hat{7}\hat{8}\hat{9}\hat{4}\hat{1}})\,,
                   &\qquad  n =1  \\
                   \\
                   \frac{\mu}{12}\Gamma^{\hat{+}\hat{3}}
                  (1-\Gamma^{\hat{7}\hat{8}\hat{9}\hat{3}\hat{2}})\,,
                   &  \qquad n =2  \\ \\
                    \frac{\mu}{12}\Gamma^{\hat{+}\hat{2}}
                  (1+\Gamma^{\hat{7}\hat{8}\hat{9}\hat{3}\hat{2}})\,,
                    & \qquad n =3  \\ \\
                      \frac{\mu}{12}\Gamma^{\hat{+}\hat{1}}
                  (1+\Gamma^{\hat{7}\hat{8}\hat{9}\hat{4}\hat{1}})\,,
                    & \qquad n =4  \\ \\
                      \frac{\mu}{24}\Gamma^{\hat{+}}
                  (\Gamma^{\hat{n}}\Gamma^{\hat{7}\hat{8}\hat{9}}
                   +3\Gamma^{\hat{7}\hat{8}\hat{9}}\Gamma^{\hat{n}})\,,
                    & \qquad  n =5, 6, 7, 8, 9.
                     \end{array} \right.
                  \label{omegan}   \eea
%%%%%

The unbroken supersymmetries of the M2-brane probe are given by the Killing spinors satisfying~\cite{Bergshoeff:1997kr}
\be \Gamma\eta =\eta\,. \label{GEE}\ee
In particular, in our supergravity setup, we consider a single  M2-brane which spans the  $(t,x^1,x^2)$
directions while being located at the origin of the transverse coordinates, $x^L=0$, $L=3,  \cdots, 10$.
Taking the static gauge,
\be
 t=\xi^0\,,\qquad  x^1=\xi^1\,, \qquad x^2=\xi^2\,, \label{static}
\ee
the projection matrix, $\Gamma$, becomes
\be \Gamma = \Gamma_{\hat{t}\hat{1}\hat{2}}\,, \ee
so that  the unbroken supersymmetry condition,  $\Gamma\eta=\eta$, reduces to
\be
\ba{cccc} \Gamma_{\hat{t}\hat{1}\hat{2}}\eta_0 = \eta_0\,,~~ &~~
 [\Gamma_{\hat{t}\hat{1}\hat{2}},\, \Omega_+ ]\eta_0=0\,,~~&~~
 [\Gamma_{\hat{t}\hat{1}\hat{2}},\, \Omega_1 ]\eta_0 =0\,,~~&~~
  [\Gamma_{\hat{t}\hat{1}\hat{2}},\, \Omega_2 ]\eta_0 =0\,.
 \ea\ee
From the explicit form of $\Omega_+,\Omega_{1,2}$, these conditions are,   at last,  equivalent to
\be\ba{lll} \Gamma_{\hat{t}\hat{1}\hat{2}}\eta_0 = \Gamma^{\hat{7}\hat{8}\hat{9}\hat{4}\hat{1}}\eta_0
=\Gamma^{\hat{7}\hat{8}\hat{9}\hat{3}\hat{2}}\eta_0 = \eta_0\,, \label{proj1} \ea\ee
which shows  that the probe configuration  preserves  four supersymmetries.
\newpage

%%%%%%%%%%%%%%%%%%%%%%%%%%%%%%%%%%%%%%%%%%%%%%%%%%%%%%%%%%%%%%%%%%%%
%%%%%%%%%%%%%%%%%%%%%%%%%%%%%%%%%%%%%%%%%%%%%%%%%%%%%%%%%%%%%%%%%%%%

\section{Derivation of the 3D massive super Yang-Mills action\label{derivation}}
In this  section, we derive explicitly the novel three dimensional massive $\N=2$ super Yang-Mills action,
(\ref{M2action}),  in three different  ways, one from the matrix model  and the other two from the
supermembrane action and the  D2-brane Dirac-Born-Infeld action. In the matrix model  setup, we derive the
full non-Abelian action, while in the M2 and D2 setups, we identify the Abelian part. All the results we
obtain here are consistent.

\subsection{Matrix model derivation}
By   expanding  the  matrix model around the above supersymmetric coincident $N$ membranes,  we derive the  massive  super Yang-Mills action. To do
so, we introduce  the gauge fields as the longitudinal fluctuations around the membranes, and write
\begin{equation}
\ba{ll}X_{i}=i\hpartial_{i}+A_{i}\,,~~~&~~i=1,2\,,\\
{}&{}\\
X_{a}=\phi_{a}\,,~~~&~~a=3,4,5,6,7,8,9\,.\ea
\end{equation}
Consequently
\begin{equation}
\ba{ll} D_{t}X_{i}=F_{0i}\,,~~~~&~~~~
{}[X_{1},X_{2}]=i(F_{12}+\theta^{-1})\,,\\
{}&{}\\
{}[X_{i},\phi]=iD_{i}\phi\,,~~~~&~~~~
{}[X_{i},\psi]=iD_{i}\psi\,, \ea
\end{equation}
where $F_{\mu\nu}=\partial_{\mu}A_{\nu}-\partial_{\nu}A_{\mu}-i[A_{\mu},A_{\nu}]$, $~\mu,\nu=0,1,2$ and  the derivative of a function along the
non-commutative coordinate is, from (\ref{non-com}), $\partial_{i}\phi=[\hpartial_{i},\phi]$.  The fields have the standard gauge transformation
properties,
\begin{equation}
\ba{ll}A_{\mu}~\rightarrow~UA_{\mu}U^{\dagger}+iU\partial_{\mu}U^{\dagger}\,,~~~~&~~~~\phi~\rightarrow~U\phi
U^{\dagger}\,.\ea
\end{equation}

To write the matrix model, (\ref{Maction}),  in terms of the gauge fields, we first note
\begin{equation}
J^{lm}\Tr(X_{l}D_{t}X_{m})= -2\Tr(\phi_{4}F_{01}+\phi_{3}F_{02}) +\displaystyle{\frac{d\,}{dt}\Tr(X_{1}X_{4}+X_{2}X_{3})\,,}~~~~l,m=1,2,3,4\,.
\end{equation}

Utilizing  the fact that the trace over the Hilbert space, ${\cal H}$, can decompose into  the  integration over the non-commutative plane and the
trace over the ``$\mbox{U}(N)$'' indices,
\begin{equation}
\displaystyle{\Tr{\cal O}(x)=\frac{1}{2\pi\theta}\int
dx^{2}\,\tr_{\!{\scriptscriptstyle N}}{\cal O}(x)\,,}
\end{equation}
one can rewrite the matrix model as a non-commutative action.  After  discarding the total
 derivative terms and the mass of the membranes, our matrix model,
 (\ref{Maction}), in the membrane  background leads to  a
 non-commutative  massive super Yang-Mills,
\begin{equation}
{\cal S}=\displaystyle{\frac{l_{p}^{6}}{~2\pi\theta R^{3}} \int\,dx^{3}}~
 \L_{0}+\mu\L_{1}+\mu^{2}\L_{2}\,, \label{ncM2action}
\end{equation}

\begin{equation}
\ba{l}{\L_{0}=\tr_{\!{\scriptscriptstyle N}}\Big[-\quarter F_{\mu\nu}F^{\mu\nu}-\half D_{\mu}\phi_{a}D^{\mu}\phi_{a}+\quarter
[\phi_{a},\phi_{b}]^{2}-i\half\psi^{\dagger}\gamma^{\mu}D_{\mu}\psi-\half\psi^{\dagger}\gamma^{a}[\phi_{a},\psi]\Big]\,,}\\
{}\\
\L_{1}=\tr_{\!{\scriptscriptstyle N}}\Big[\textstyle{\frac{1}{3}}(\phi_{4}F_{01}+\phi_{3}F_{02})
+\textstyle{\frac{1}{6}}\epsilon^{pq}\phi_{p}D_{0}\phi_{q}
-i\textstyle{\frac{1}{3}}\epsilon^{rst}\phi_{r}\phi_{s}\phi_{t}+i\textstyle{\frac{1}{24}}\psi^{\dagger}\Pi\psi \Big]\,,\\
{}\\
{\L_{2}=-\,\half(\textstyle{\frac{1}{3}})^{2}\,\tr_{\!{\scriptscriptstyle N}}\left(\phi_{7}^{\,2}+\phi_{8}^{\,2}+\phi_{9}^{\,2}\right)\,,}
\ea\label{ncM2LLL}
\end{equation}
where $\mu=0,1,2$, $a=3,4,5,6,7,8,9$, $p=5,6$, $r=7,8,9$, $\epsilon^{56}=\epsilon^{789}=1$,  and
\begin{equation}
\Pi=(\gamma^{14}+\gamma^{23}-\gamma^{56}+3\gamma^{789})\,.
\end{equation}
Any product is to  be understood as   the non-commutative star product.
%%%
%%The coupling constant is given by
%%\begin{equation}
%%\displaystyle{g^{2}=\frac{2\pi\theta R^{3}}{l_{p}^{6}}}\,.
%%%
%% =\frac{2\pi\theta g_s}{l_{s}^{3}}}\,.
%%%
%%\label{g2theta}
%%\end{equation}
%%%%
The dynamical supersymmetry transformations are, from (\ref{susytr}),
\begin{equation}
\ba{l} \delta A_{\mu}=i\psi^{\dagger}\gamma_{\mu}\E(t)\,,~~~~~~~~~~~\delta \phi_{a}=i\psi^{\dagger}\gamma_{a}\E(t)\,,\\
{}\\
\delta\psi=\left[\ba{l}\half F_{\mu\nu}\tilde{\gamma}^{\mu}\gamma^{\nu}+D_{\mu}\phi_{a}\tilde{\gamma}^{\mu}\gamma^{a}-i\half
[\phi_{a},\phi_{b}]\gamma^{ab}-\textstyle{\frac{\mu}{3}}(\phi_{7}\gamma^{7}+\phi_{8}\gamma^{8}+\phi_{9}\gamma^{9})\gamma^{789}\\
{}\\
+\frac{1}{\theta}\gamma^{12}+\textstyle{\frac{\mu}{6}}\left((-\frac{1}{\theta}x^{2}+A_{1})\gamma^{1}
+\phi_{4}\gamma^{4}\right)(\gamma^{789}-\gamma^{14})\\
{}\\
+\textstyle{\frac{\mu}{6}}\left((\frac{1}{\theta}x^{1}+A_{2})\gamma^{2}
+\phi_{3}\gamma^{3}\right)(\gamma^{789}-\gamma^{23})+\textstyle{\frac{\mu}{6}}(\phi_{5}\gamma^{5}+\phi_{6}\gamma^{6})
(\gamma^{789}+\gamma^{56})\ea\right]\E(t)\,, \label{M2susytr}\ea
\end{equation}
where $a=3,4,5,6,7,8,9$ and
\begin{equation}
\ba{cc} \displaystyle{\E(t)=e^{\frac{\mu}{12}(-\gamma^{14}-\gamma^{23}+\gamma^{56}+\gamma^{789})t}\E\,,}~~~&~~~ \E={C}\E^{\ast}\,.\ea\label{ncEt}
\end{equation}
Note that the full supersymmetry remains unbroken for this reformulation, which is no surprise as the
non-commutative three dimensional action (\ref{ncM2action}) is merely  a particular manifestation of the
background independent
M-theory matrix model \cite{Seiberg:2000zk,JHPcomments}. \\

Despite the similarity between the terms mixing the field strength with the Higgs and the Chern-Simons term,
there is no quantization rule for the coefficient, contrary to the Chern-Simons theory  on a non-commutative
plane \cite{NC-CS}, since the terms here  are manifestly gauge
invariant.\\

By taking the commutative limit, $\theta\rightarrow 0$ while keeping the coupling constant, $2\pi\theta R^{3}/l_{p}^{6}$,  fixed, one can obtain a
commutative  action, which is exactly of the same form as (\ref{M2action}), but $\mu$ therein is replaced by $\sqrt{2}\mu$. This $\sqrt{2}$ factor
can be absorbed by scaling the worldvolume coordinates and redefining the field variables as
\begin{equation}
(x^{\mu},~A_{\mu},~\phi_{a},~\psi)~~\longrightarrow ~~(2^{-1/2}x^{\mu},~2^{1/2}A_{\mu},~2^{1/2}\phi_{a},~2^{3/4}\psi)\,.\label{change}
\end{equation}
This scaling will be justified in the subsection \ref{compare}.\\

In the commutative limit, some supersymmetries become singular and broken. In the  subsection \ref{Matrixsusy}, we will show that only four
supersymmetries survive.

\subsection{Derivation from the supermembrane action}
%%%%%%%%%%%%%%%%%%%%%%%%%%%%%%%%%%%%%%%%%%%%%%%%%%%%%%%%%%%%%%%%%%%%

In this subsection,  we obtain the  low energy  effective worldvolume action for the  membrane spanning the
$(x^{0},x^{1},x^{2})$ directions from the supermembrane action. What we mean by ``low energy" is the
following limits. We scale the M2-brane tension as $T_{M2}=1/(4\pi^2l_p^3)\sim\epsilon^{-2}\rightarrow
\infty$, and let the transverse coordinates $x^l,y=x^{10},\, l=3,...,9$, the fermionic superpartner,
$\theta$, scale like $(x^l,y,\theta) \sim \epsilon$. This, after the compactification along the $y$
direction, corresponds to the scaling of the string length and the string coupling as $l_s\sim g_s\sim
\epsilon^{1/2}\rightarrow 0$,  keeping $\g^2=g_s/l_s$ finite. \\

In the above  scaling limits,  ${\cal M}^2={\cal O}(\theta^2)\sim \epsilon^2$ and,  from (\ref{E1}) and
(\ref{E2}),  \cite{deWitpp,deWit:1998yu}
\bea && E^{\hat{R}}_\alpha = -(\bar{\theta}\Gamma^{\hat{R}})_\alpha
+{\cal O}(\epsilon^3) \, , \\
\nonumber \\
&&E^{\hat{R}}_M = e^{\hat{R}}_M + \bar{\theta}\Gamma^{\hat{R}}
\Big(-\frac{1}{4}\omega^{\hat{R}\hat{S}}_M\Gamma_{\hat{R}\hat{S}} +
T_M^{NPQR} F_{NPQR}
\Big)\theta + {\cal O}(\epsilon^3) \, , \\
\nonumber \\
&& E^{\hat{a}}_\alpha = \delta^{\hat{a}}_\alpha + {\cal
O}(\epsilon^2)  \,, \\ \nonumber \\
&& E^{\hat{a}}_M =\Big[
\Big(-\frac{1}{4}\omega^{\hat{R}\hat{S}}_M\Gamma_{\hat{R}\hat{S}} +
T_M^{NPQR} F_{NPQR} \Big)\theta \Big]^{\hat{a}}  + {\cal
O}(\epsilon^2) \,. \eea
Thus,
\bea  \Pi^{~\hat{R}}_\mu &=&
\partial_\mu x^M e^{\hat{R}}_M +
\partial_\mu x^M\bar{\theta}\Gamma^{\hat{R}}
\Big(-\frac{1}{4}\omega^{\hat{S}\hat{U}}_M\Gamma_{\hat{S}\hat{U}} +
T_M^{NPQR} F_{NPQR} \Big)\theta +
\bar{\theta}\Gamma^{\hat{R}}\partial_\mu\theta +{\cal
O}(\epsilon^3) \,, \nonumber \\             {}\\
\Pi^{~\hat{a}}_\mu &=&  \partial_\mu \theta^{\hat{a}} +
\partial_\mu x^M
\Big[\Big(-\frac{1}{4}\omega^{\hat{R}\hat{S}}_M\Gamma_{\hat{R}\hat{S}}
+ T_M^{NPQR} F_{NPQR} \Big)\theta\Big]^{\hat{a}} + {\cal
  O}(\epsilon^2)\,. \nonumber\eea
Adopting the static gauge (\ref{static}) and letting
\be
\Omega_{t}=\Omega_{0}\equiv\textstyle{\frac{1}{\sqrt{2}}}\Omega_{+} =
\textstyle{\frac{\mu}{12\sqrt{2}}}
(\Gamma^{\hat{+}\hat{-}\hat{7}\hat{8}\hat{9}}
         +\Gamma^{\hat{1}\hat{4}}
  +\Gamma^{\hat{2}\hat{3}}-2\Gamma^{\hat{7}\hat{8}\hat{9}})\,,
\label{Omegat}\ee
the three form superfield becomes\footnote{Note that the first term in
  (\ref{Bform}) scales as $\epsilon^{3}$ in our setup.}
\be B =d\xi^{\mu}\wedge d\xi^{\nu} \wedge d\xi^{\rho}\Big(\frac{1}{2}
 \bar{\theta} \Gamma_{\hat{\mu}\hat{\nu}} {D}_{\rho}\theta \Big)
 + {\cal O}(\epsilon^3) \,,\ee
while  the worldvolume induced metric is explicitly, with $l=3,\cdots,9$,
\bea  h_{00} &=&-(1+\tilde{H}/2)+(\partial_0y)^2+(\partial_0x^l)^2
-2\bar{\theta}\Gamma^{\hat{t}}(\partial_0\theta + \Omega_t\theta)
+ {\cal O}(\epsilon^3) \,,\nonumber \\
\nonumber \\
 h_{11} &=& 1 + (\partial_1 y)^2 + (\partial_1x^l)^2 +
\frac{2\mu}{3\sqrt{2}}x^4\partial_1 y  +
2\bar{\theta}\Gamma^{\hat{1}}(\partial_1\theta + \Omega_1\theta)
+ {\cal O}(\epsilon^3) \,, \nonumber\\
 \nonumber\\
 h_{22} &=& 1 +
(\partial_2 y)^2 + (\partial_2x^l)^2
+\frac{2\mu}{3\sqrt{2}}x^3\partial_2 y +
2\bar{\theta}\Gamma^{\hat{2}}(\partial_2\theta + \Omega_2\theta)
+ {\cal O}(\epsilon^3) \, ,\nonumber \\
\nonumber \\
 h_{01} & =& \frac{\mu}{3\sqrt{2}}x^4 +{\cal
O}(\epsilon^2) \,, \qquad h_{02} = \frac{\mu}{3\sqrt{2}}x^3 +{\cal O}(\epsilon^2) \,, \qquad h_{12} = {\cal O}(\epsilon^2) \,. \eea
Hence,
\be -\det h_{\mu\nu} = 1 + \frac{H}{2} + \partial_\mu y\partial^{\mu}
y +\partial_\mu x^l\partial^{\mu}x^l +
\frac{2\mu}{3\sqrt{2}}(x^4\partial_1 y
+ x^3\partial_2 y)
+2\bar{\theta}\Gamma^{\hat{\mu}}(\partial_\mu+\Omega_{\mu})\theta+
{\cal O}(\epsilon^3)\,.\ee
In the above,  we have introduced
\be\textstyle{\tilde{H}= H  -(\frac{\mu}{3})^{2}(x_3^2+x_4^2)=
(\frac{\mu}{6})^{2}(x_5^2+x_6^2)+(\frac{\mu}{3})^{2}(x_{7}^{2}+x_8^2+x_9^2)
\,.}\ee

After fixing  the $\kappa$-symmetry as\footnote{An identical gauge
choice in the string case was considered in~\cite{Cvetic:2002nh}
and called ``physical gauge''. }
\be (1+\Gamma_{\hat{t}\hat{1}\hat{2}})\theta=0\,, \label{theta} \ee
the Wess-Zumino term becomes
\be\int B = -\int d\xi^{3}\,
\bar{\theta}\Gamma^{\hat{\mu}}(\partial_\mu+\Omega_\mu)\theta + {\cal
  O}(\epsilon^3)\,.\ee
Writing  the 11D gamma matrices  in terms of the $16\times 16$ Euclidean 9D gamma matrices\footnote{Here, for
  simplicity, we
drop the hat symbol  for the flat spacetime index in the  9D gamma matrix.} and $\gamma^{0}=-1$, (\ref{11Dgamma}),  the $\kappa$-symmetry fixing
condition, (\ref{theta}), can be solved by a 16-component 9D Majorana spinor, $\psi$,
\bea \theta = \frac{1}{2\sqrt{2}}{\psi\choose -\gamma^{12}\psi} \,,
\qquad \bar{\theta} = \frac{i}{2\sqrt{2}}(-\psi^\dagger\gamma^{12}\,,
\psi^\dagger)\,. \label{psi16}\eea
To express the final form of the action in terms of the finite quantities, we replace the  transverse coordinates,
\be (x^l,\, y,\,\psi )\longrightarrow 2\pi l_{s}^{2}(\phi^l,\,
\phi^{y},\,\psi)\,,\qquad l=3,\cdots,9\,. \label{conversion}\ee
Now, in the low energy limit,  the supermembrane action  reduces to, with $L=3,\cdots,9,y$,
\begin{equation}
\ba{ll} S_{M2} =&\displaystyle{ -T_{M2}\int {d}\xi^{3}~ 1}\\
{}&{}\\
{}& +\displaystyle{\frac{1}{{(g_s/l_s)}}}\displaystyle{\int{d\xi^{3}}}
\left[\ba{l}-\half\partial_{\mu}\phi^{L}\partial^{\mu}\phi^{L}-
  i\half\psi^\dagger\gamma^{\mu}\partial_\mu\psi\\
{}\\
-\textstyle{\frac{\mu}{3\sqrt{2}}(\phi^4\partial_{1}\phi^{y} +
  \phi^3\partial_{2}\phi^{y})+
i\frac{\mu}{24\sqrt{2}}\psi^\dagger(\gamma^{24}-\gamma^{13}+
3\gamma^{789})\psi}\\
{}\\
-\half\textstyle{(\frac{\mu}{6\sqrt{2}})^{2}(\phi_5^2+\phi_6^2)-
\half(\frac{\mu}{3\sqrt{2}})^{2}(\phi_{3}^2+\phi_4^2+\phi_7^2+\phi_8^2+\phi_9^2)}
\ea
 \right] \,.\ea\label{SM2}
 \end{equation}
The dualization of this action to a $\mbox{U}(1)$ gauge theory  is performed in the subsection \ref{compare}.

%%%%%%%%%%%%%%%%%%%%%%%%%%%%%%%%%%%%%%%%%%%%%%%%%%%%%%%%%%%%%%%%%%%%%%%%%%%%%%%%%%%%%%%%%%%%%%%%%%%%%%%%%%%%%%%%%%%%%%%%%%%%%%%%%%%%%%%%%%

\subsection{Derivation from the D2 Dirac-Born-Infeld action}
In this subsection, we derive the  massive gauge theory as a low energy limit  of the Dirac-Born-Infeld action.  As the explicit form of the
supersymmetric DBI action in terms of the component fields is not known in the generic background or the pp-wave background\footnote{For the
superfield formalism, see \cite{Bergshoeff:1996tu,Aganagic:1997zk}.},
we focus on the bosonic sector.\\

Writing the eleven dimensional pp-wave geometry, (\ref{metric}),  as
\be ds^2_{11} = e^{-2\phi/3}ds^2_{IIA}+e^{4\phi/3}(dy+C_{(1)})^2\,,  \ee
and compactifying the $y$ direction, we obtain the ten dimensional type IIA supergravity background,
\bea ds^2_{IIA}&=&-(1-\tilde{H}/2)^{-1/2}\Big( dt-\frac{\mu}{3\sqrt{2}}x^4dx^1
          -\frac{\mu}{3\sqrt{2}}x^3dx^2\Big)^2
 +(1-\tilde{H}/2)^{1/2}\sum_{n=1}^9dx^ndx^n\,, \nonumber \\ \nonumber\\
 e^\phi  &=& (1-\tilde{H}/2)^{3/4} \,.
\eea
In the limit, $x^{M}\sim \epsilon\rightarrow 0,\, M=3,4,\cdots$, we have been taking, the dilaton is real and small.  The non-vanishing components of
the RR one form, $C_{(1)}=C_M dx^M$, are
\bea
 C_t &=&-\frac{\tilde{H}}{2}(1-\tilde{H}/2)^{-1} = {\cal O}(\epsilon^2)\, , \nonumber \\
\nonumber \\
 C_1 &=&\frac{\mu}{3\sqrt{2}}x^4(1-\tilde{H}/2)^{-1}
=\frac{\mu}{3\sqrt{2}}x^4 +{\cal O}(\epsilon^2)\,, \label{C1} \\
\nonumber \\
  C_2 &=&\frac{\mu}{3\sqrt{2}}x^3(1-\tilde{H}/2)^{-1}
=\frac{\mu}{3\sqrt{2}}x^3+{\cal O}(\epsilon^2)\,, \nonumber\eea
while the three form and four form fluxes  are
\begin{equation}
\ba{cc}H_{789}=(dB^{IIA})_{789}=\frac{\mu}{\sqrt{2}}\,,~~~&~~~~
F_{t789}=(dC_{(3)})_{t789}=\frac{\mu}{\sqrt{2}}\,.\ea
\end{equation}
A few comments are in order.  The resulting 10D background breaks all the  supersymmetries  in the type  IIA
supergravity, since no constraint on the constant spinor, $\eta_{0}$, removes the $y$ dependence from   the
11D Killing spinor expression,  (\ref{Killing}).  From (\ref{O+}),  the periodic identification  over  the
$y$ direction is compatible with the  Killing spinors, (\ref{Killing}),  only for the special values of the
compactification radii, e.g.  zero  \cite{Michelson:2002wa}.\\

The  action describing the D2-brane consists of the  Dirac-Born-Infeld and the  Wess-Zumino terms
\cite{Cederwall:1996ri},
\begin{equation}
\ba{c} S_{D2} = S_{DBI} + S_{WZ}\,,\\
{}\\
S_{DBI} = -T_{D2}\displaystyle{\int d\xi^{3}}\,
e^{-\phi}\sqrt{-\det(h_{\mu\nu} + {\cal F}_{\mu\nu})}\,, \\
{}\\
 S_{WZ}  = T_{D2}\displaystyle{\int} \Big(C_{(1)}\wedge {\cal F} +
C_{(3)}\Big)\,, \ea
\end{equation}
where  $T_{D2}=1/(4\pi^2g_sl_s^3)$, and
\begin{equation}
\ba{cc}h_{\mu\nu} = \partial_\mu x^M\partial_{\nu}x^N
g_{MN}^{IIA}\,,~~~&~~~{\cal F}_{\mu\nu} = 2\pi l^2_s F_{\mu\nu} -
\partial_\mu x^M\partial_{\nu}x^N B_{MN}^{IIA}\,.\ea\label{2pil}
\end{equation}
\\

Now adopting the static gauge, (\ref{static}), replacing  the transverse
coordinates, $x^l$, by $2\pi l_s^2 \phi^l$,    $l=3,\cdots,9$, and taking
the limit, $l_{s}^{2}\sim\epsilon\rightarrow 0$ while keeping
$\g^{2}=g_{s}/l_{s}$ finite, the terms involving $B^{IIA}$ and $C_{(3)}$
vanish. In this low energy limit, the above D2-brane action becomes
\be\ba{ll} S_{D2}=& -T_{D2}\displaystyle{\int d\xi^{3}~ 1}\\
{}&{}\\
{}&\displaystyle{+\frac{1}{(g_{s}/l_{s})}\int d\xi^{3}}\left[\ba{l}
  -\quarter F_{\mu\nu}F^{\mu\nu} -\half\partial_\mu
  \phi^l\partial^{\mu}\phi^{l}
+\textstyle{\frac{\mu}{3\sqrt{2}}}(\phi^4 F_{02} - \phi^3 F_{01} )\\
{}\\
-\half\textstyle{(\frac{\mu}{6\sqrt{2}})^{2}(\phi_5^2+\phi_6^2)-
  \half(\frac{\mu}{3\sqrt{2}})^{2}(\phi_7^2+\phi_8^2+\phi_9^2)}
\ea\right]\,.\ea\label{SD2}\ee

Although the precise form of the non-Abelian Dirac-Born-Infeld action is not known ({\textit{cf.}
\cite{NDBI}), the non-Abelian generalization of the above bosonic quadratic action can be done following the
Myers' prescription  \cite{Myers:1999ps}, which will result in the bosonic part of
(\ref{M2action}).\footnote{However, in general, there is  an ambiguity when one  tries to do the non-Abelian
generalization.   One can  put an arbitrary numerical factor, say $\lambda$,  in front of any commutator.
The appropriate  scaling of the fields like $A_{\mu}\rightarrow A_{\mu}/\lambda$,  may absorb the numerical
factor, but alters the string length in (\ref{2pil}) as $l_{s}\rightarrow l_{s}/\sqrt{\lambda}$. Hence,
different choices are physically distinct. Unfortunately, we are not able to fix the value in our framework,
but set $\lambda=1$ in (\ref{M2action}) for simplicity.} Contrary to the supermembrane case, all the terms
linear in $\mu$, including the Myers term, arise from the Wess-Zumino term.

%%%%%%%%%%%%%%%%%%%%%%%%%%%%%%%%%%%%%%%%%%%%%%%%%%%%%%%%%%%%%%%%%%%%%%%%%%%%%%%%%%%%
%%%%%%%%%%%%%%%%%%%%%%%%%%%%%%%%%%%%%%%%%%%%%%%%%%%%%%%%%%%%%%%%%%%%%%%%%%%%%%%%%%%%%%%%%%%%%%%%%%%%%%%%%%%%
%%%%%%%%%%%%%%%%%%%%%%%%%%%%%%%%%%%%%%%%%%%%%%%%%%%%%%%%%%%%%%%%%%%%%%%%%%%%%%%%%%%%%%%%%%%%%%%%%%%%%%%%%%%%
\subsection{Mutual agreement among the results through the dualization \label{compare}}
In this subsection, we compare the resulting three actions, ${\cal S}$ from the matrix model
(\ref{M2action}), $S_{M2}$ from the supermembrane (\ref{SM2}),  and $S_{D2}$ from the D2-brane (\ref{SD2}).
By tuning the gauge choices in each setup to the consistent one, we show that all the actions agree with
another.\\

Before starting, we justify the scaling, (\ref{change}),  we took in the last step of the derivation of the
action, ${\cal S}$, in the matrix model setup. The scaling of the field variables is  merely a field
redefinition, while  that  of the worldvolume coordinates is taken  to make the choice of the ``time"
coordinate in the matrix model consistent  with the static gauge in the M2/D2 action,
\be\ba{ll}\xi^{0}= x^{+}~~\longrightarrow~~\xi^{0}=\sqrt{2}x^{+}= x^{0}+x^{10}~\sim~x^{0}\,,\ea\ee
since the compactification radius, $R_{y}$,  is vanishingly small and   $0\leq y=x^{10}<2\pi R_{y}\rightarrow 0$.\\

It is  worth to note that, although  the periodic identification  over   the $y$ direction is not compatible
with the 11D Killing spinors for the generic values of the  radii, in the  small radius limit, the
compactified  pp-wave geometry may well recover the full supersymmetries. One way to understand this is
going back to the light-cone coordinates, $x^\pm=(t\pm y)/\sqrt{2}$, which are periodic as $x^\pm\sim x^\pm
\pm 2\pi(R_y/\sqrt{2})$.  We  take the infinite boost along the $y$ direction such that the compactification
over the $y$ direction turns into  that over the $x^-$ direction  of a finite radius. In the limit, the
$x^+$ coordinate possesses no periodicity  and serves the role of the ``time" coordinate. Since the 11D
Killing spinors,  (\ref{Killing}),  are independent of  $x^-$, no supersymmetry is broken under the
compactification over the $x^{-}$ direction.  In fact,  this  was the basic setup  the M-theory matrix model
was originally obtained \cite{Banks:1996vh,Susskind:1997cw,Sen,Seiberg}. We also note that, in the same
limit, the supersymmetric membrane configuration spanning  the $(x^{0},x^{1},x^{2})$ directions in the probe
analysis can be identified with the one  spanning the $(x^{+},x^{1},x^{2})$ directions in the  matrix model.\\

Now we identify $S_{M2}$ with $S_{D2}$ through the  dualization of the gauge fields to the compact
scalar\footnote{Note that, from $2\pi l_{s}^{2}\phi^{y}=y$ and $R_{y}=g_{s}l_{s}$,  in the low energy limit,
$g_{s}\sim\l_{s}\sim\epsilon^{1/2}\rightarrow 0$, the periodicity of $\phi^{y}$ is  finite, $g_s/l_s$. },
$\phi^{y}$. We add a total derivative term to $S_{M2}$,
\begin{equation}
S_{M2}~\longrightarrow~S_{M2}+\displaystyle{\frac{1}{{(g_s/l_s)}}}\displaystyle{\int d\xi^{3}}~
\epsilon^{\mu\nu\rho}\partial_{\mu}\phi^{y}\partial_{\nu}A_{\rho}\,,\label{Stotal}
\end{equation}
and integrate out the  scalar.   Effectively this replaces the derivatives of $\phi^{y}$ in the right hand
side  of (\ref{Stotal})  by
\begin{equation}
\ba{ccc}
\partial_{0}\phi^{y}=-F_{12}\,,~~~&~~~\partial_{1}\phi^{y}=F_{20}-\textstyle{\frac{\mu}{3\sqrt{2}}}\phi^{4}\,,
~~~&~~~\partial_{2}\phi^{y}=F_{01}-\textstyle{\frac{\mu}{3\sqrt{2}}}\phi^{3}\,, \ea\label{dual}
\end{equation}
which results in the supersymmetric completion of $S_{D2}$,
\begin{equation}
{\cal S}^{\prime}=\displaystyle{\frac{1}{{(g_s/l_s)}}}\displaystyle{\int d\xi^{3}}
\left[\ba{l} -\quarter F_{\mu\nu}F^{\mu\nu}-\half\partial_{\mu}\phi^{l}\partial^{\mu}\phi^{l}-i\half\psi^\dagger\gamma^{\mu}\partial_\mu\psi\\
{}\\
+\textstyle{\frac{\mu}{3\sqrt{2}}}(\phi^4 F_{02} - \phi^3 F_{01} )+i\frac{\mu}{24\sqrt{2}}\psi^\dagger(\gamma^{24}-\gamma^{13}+3\gamma^{789})\psi\\
{}\\
-\quarter\textstyle{(\frac{\mu}{6})^{2}(\phi_5^2+\phi_6^2)-\quarter(\frac{\mu}{3})^{2}(\phi_7^2+\phi_8^2+\phi_9^2)}\ea
 \right]\,,\label{calSD2}
\end{equation}
where $l=3,4,\cdots,9$.\\

Finally,  we match ${\cal S}^{\prime}$  with   the Abelian version of ${\cal S}$. As done in the matrix model
setup, (\ref{rota}) and (\ref{PSIROTATION}), we rotate the scalars and the fermion, $\phi^5, \phi^6$, $\psi$
in ${\cal S}^{\prime}$ such  that the mass terms for the scalars disappear and that for the fermion gets
modified. As stated earlier, this removes the explicit time dependency in the worldvolume supersymmetry
transformations. The resulting action is of  the same form as ${\cal S}$, except the  $\pi/2$ rotation of
the worldvolume coordinates,
\begin{equation}
(\xi^{0},\,\xi^{1},\,\xi^{2})\longrightarrow (\xi^{0},\,-\xi^{2},\,\xi^{1})\,,\label{gchange}
\end{equation}
which accompanies  $(\gamma^{1},\gamma^{2})\rightarrow(-\gamma^{2},\gamma^{1})$.  This  $\pi/2$ rotation is
an artifact  of the two different gauge choices taken in the matrix model  and in the supergravity analysis,
since, in the matrix model setup, we choose the worldvolume coordinates as $X^{1}=-\xi^{2}/\theta$,
$X^{2}=\xi^{1}/\theta$, (\ref{M2sol}) and (\ref{M2sol2}). After the rotation, ${\cal S}^{\prime}$, is exactly
mapped to the Abelian part of ${\cal S}$.\newpage

%%%%%%%%%%%%%%%%%%%%%%%%%%%%%%%%%%%%%%%%%%%%%%%%%%%%%%%%%%%%%%%%%%%%%%%%%%%%%%%%%%%%%%%%%%%%%%%%%%%%%%%%%%%%
%%%%%%%%%%%%%%%%%%%%%%%%%%%%%%%%%%%%%%%%%%%%%%%%%%%%%%%%%%%%%%%%%%%%%%%%%%%%%%%%%%%%%%%%%%%%%%%%%%%%%%%%%%%%

%%%%%%%%%%%%%%%%%%%%%%%%%%%%%%%%%%%%%%%%%%%%%%%%%%%%%%%%%%%%%%%%%%%%%%%%%%%%%%%%%%%%%%%%%%%%%%%%%%%%%%%%%%%%
%%%%%%%%%%%%%%%%%%%%%%%%%%%%%%%%%%%%%%%%%%%%%%%%%%%%%%%%%%%%%%%%%%%%%%%%%%%%%%%%%%%%%%%%%%%%%%%%%%%%%%%%%%%%
\section{Worldvolume supersymmetry\label{wsusy}}
In this section, we derive the worldvolume supersymmetries from the matrix model and from the supergravity,
respectively.
\subsection{Worldvolume SUSY from the matrix model\label{Matrixsusy}}
As seen in (\ref{M2susytr}), the dynamical supersymmetry transformation of the fermions becomes singular
when we take the commutative limit, $\theta\rightarrow 0$. To remedy the problem, one should first impose
the following   constraints on the Killing spinors,
\begin{equation}
\gamma^{14}\E=\gamma^{23}\E=\gamma^{789}\E\,,\label{EEG}
\end{equation}
which in turn implies $\gamma^{56}\E=\gamma^{789}\E$, so that    the time dependency of $\E(t)$,
(\ref{ncEt}), effectively disappears. The only remaining singular part is now $(1/\theta)\gamma_{12}\E$, and
this can be  removed  by adding the kinematical supersymmetry given  by
$\E^{\prime}=-(1/\theta)\gamma_{12}\E$. Again the time dependency of the kinematical supersymmetry
transformation drops out. Therefore, the unbroken supersymmetries of the membranes reappear precisely  as
the supersymmetry of the worldvolume theory. As stated before, the whole constraints on the Killing spinors,
(\ref{EEG}), can be rewritten  in a concise manner, using the  projection matrix  (\ref{Omega}),
\begin{equation}
\Omega\E=\E\,.\label{OmegaEE}
\end{equation}
The worldvolume supersymmetry transformations of the action, ${\cal S}$,  (\ref{M2action}),  are then
\begin{equation}
\ba{l}\delta A_{\mu}=i\psi^{\dagger}\gamma_{\mu}\E\,,~~~~~~~~~~~
\delta \phi_{a}=i\psi^{\dagger}\gamma_{a}\E\,,\\
{}\\
\delta\psi=\Big[\half F_{\mu\nu}\tilde{\gamma}^{\mu}\gamma^{\nu}+D_{\mu}\phi_{a}\tilde{\gamma}^{\mu}\gamma^{a}-i\half [\phi_{a},\phi_{b}]\gamma^{ab}
+\textstyle{\frac{\mu}{3\sqrt{2}}}(\phi_{p}\gamma^{p}-\phi_{r}\gamma^{r})\gamma^{789} \Big]\E\,,\ea\label{WVsusytr}
\end{equation}
where $p=5,6$, $r=7,8,9$ and $\E$ is  a time independent constant Majorana spinor subject to (\ref{OmegaEE}).\\
%%
% Note that the supersymmetry transformations do not have the explicit time dependency so that the supercharges commute with the Hamiltonian.\\
%%

It is interesting   to compare  with the ordinary BFSS matrix model  or the $\mu=0$ case. In that case, the only singular piece in the
$\theta\rightarrow 0$ limit of  the dynamical supersymmetry transformation is $(1/\theta)\gamma_{12}\E$, and  this  can be completely removed by the
kinematical supersymmetry transformation. Thus, both in the $\mu=0$ and $\mu\neq 0$ cases, the commutative worldvolume actions possess  the
same numbers of supersymmetries the background membranes preserve, i.e. 16 for $\mu=0$, and 4 for  $\mu\neq 0$.\\

%%%%%%%%%%%%%%%%%%%%%%%%%%%%%%%%%%%%%%%%%%%%%%%%%%%%%%%%%%%%%%%%%%%%%%%%%%%
\subsection{Worldvolume SUSY from the supermembrane action\label{SSS}}
%%%%%%%%%%%%%%%%%%%%%%%%%%%%%%%%%%%%%%%%%%%%%%%%%%%%%%%%%%%%%%%%%%%%%%%%
In this subsection, we derive the worldvolume supersymmetry transformations of the quadratic actions,   $S_{M2}$ (\ref{SM2}) and  ${\cal S}^{\prime}$
(\ref{calSD2}). The worldvolume supersymmetry is identified as a specific combination of the spacetime supersymmetry and the $\kappa$-symmetry,
\be\ba{ll} \delta \theta =\eta+ (1+\Gamma)\kappa \,, ~~~&~~~ \delta x^M =
\bar{\theta}\Gamma^M\eta-\bar{\theta}\Gamma^M(1+\Gamma)\kappa\,,~~~M=0,1,\cdots,9,y\,, \ea\label{wvsusy}\ee
which must preserve   the $\kappa$-symmetry fixing, (\ref{theta}),  as well as the static gauge, (\ref{static})
\cite{deWitpp,Bergshoeff:1997kr,Cvetic:2002nh},
\bea&&(1+\Gamma_{\hat{t}\hat{1}\hat{2}})\delta\theta = 0\,,\label{1st}\\{}\nonumber\\
&&\delta x^{0}=\delta x^{1}=\delta x^{2}=0\,.\label{2nd}\eea
In the case of the pp-wave background geometry we consider,   the $\kappa$-symmetry  parameter,
$\kappa(\xi)$, is an arbitrary fermionic `local' variable, while the Killing spinor, $\eta$, is of the fixed
form, (\ref{Killing}),  with an arbitrary `constant' spinor, $\eta_{0}$. From (\ref{metric}),  there exist
translational  isometries in the $(x^{0},x^{1},x^{2})$ directions. However, these rigid isometries serve no
role to ensure the
vanishing of the local transformations, (\ref{2nd}).\\

In the limit, $(x^{L},\theta)\sim\epsilon\rightarrow 0,~L=3,\cdots,9,y$,  the Lagrangian terminates  at the quadratic order in $\epsilon$, and the
worldvolume supersymmetry transformations are to be kept up to the linear order. From (\ref{KG}), (\ref{Killing}),  (\ref{theta}),
$~~~\bar{\theta}(1+\Gamma_{\hat{t}\hat{1}\hat{2}})=0$ and
\be \ba{ll} \delta\theta=&(1+\Gamma_{\hat{t}\hat{1}\hat{2}})\kappa_{0}(\xi)+(1-\xi^{1}\Omega_{1}-\xi^{2}\Omega_{2})e^{-t\Omega_{t}}\eta_{0}\\
{}&{}\\
{}&+(1+\Gamma_{\hat{t}\hat{1}\hat{2}})\kappa_{1}(\xi)+\partial_\mu x^L\Gamma_{\hat{L}}\Gamma^{\hat{\mu}}\Gamma_{\hat{t}\hat{1}\hat{2}}\kappa_{0}(\xi)
-(\displaystyle{\sum_{n=3}^{9}x^{n}\Omega_{n}}+y\Omega_{t})e^{-t\Omega_{t}}\eta_{0}+{\cal
O}(\epsilon^{2})\,,\\
{}&{}\\
\multicolumn{2}{l}{\delta x^{\mu}=\bar{\theta}\Gamma^{\mu}(1-\xi^{1}\Omega_{1}-\xi^{2}\Omega_{2})e^{-t\Omega_{t}}\eta_{0}+{\cal O}(\epsilon^{2})
\,,} \ea \ee
where $L=3, \cdots, 9,y$ and $\kappa_{0}(\xi),\kappa_{1}(\xi)$ denote the zeroth, first order of $\kappa(\xi)$ in $\epsilon$.

Imposing the constraint,  (\ref{1st}), one can solve for $(1+\Gamma_{\hat{t}\hat{1}\hat{2}})\kappa_{0}$,
$(1+\Gamma_{\hat{t}\hat{1}\hat{2}})\kappa_{1}$, and  the vanishing of $\delta x^{\mu}$ is equivalent to
\be(1-\Gamma_{\hat{t}\hat{1}\hat{2}})(1-\xi^{1}\Omega_{1}-\xi^{2}\Omega_{2})e^{-t\Omega_{t}}\eta_{0}=0\,.\ee
This relation must hold for arbitrary   $\xi^{1},\xi^{2},t$ so that, from (\ref{omegan}), we get the same
constraints on $\eta_{0}$ as in the probe analysis, (\ref{proj1}),
\be\ba{lll} \Gamma_{\hat{t}\hat{1}\hat{2}}\eta_0 = \Gamma^{\hat{7}\hat{8}\hat{9}\hat{4}\hat{1}}\eta_0
=\Gamma^{\hat{7}\hat{8}\hat{9}\hat{3}\hat{2}}\eta_0 = \eta_0\,. \ea\ee
After all, from  (\ref{conversion}), in terms of the 16-component spinors, $\psi$ and  $\E$, of which the
latter gives the solution of $(1-\Gamma_{\hat{t}\hat{1}\hat{2}})\eta_{0}=0$,
\be \eta_0 = \textstyle{\frac{1}{\sqrt{2}}}\displaystyle{\left(\ba{c}{\cal E}\\ \gamma^{12}{\cal E}\ea\right)} \,, \ee
the worldvolume supersymmetry transformations  read
\bea &&\delta \phi^l = i\psi^\dagger \gamma^l {\cal E}(t)\,, \qquad \delta \phi^y =  -i\psi^\dagger
\gamma^{12} {\cal E}(t)\,, \qquad l=3,4,\cdots 9\,,
 \nonumber              \\ \nonumber \\
  &&\delta \psi = \left[-\partial_\mu \phi^l\gamma_l\gamma^{\mu} -
\partial_\mu
\phi^y\tilde{\gamma}^{\mu}\gamma^{12} + {\textstyle\frac{\mu}{3\sqrt{2}}} (x^3\gamma^1 - x^4\gamma^2) - {\textstyle\frac{\mu}{12\sqrt{2}}}
\phi^p(\gamma_p\gamma^{789}+3\gamma^{789}\gamma_p)
\right]{\cal E}(t) \,,               \nonumber        \\ \nonumber \\
&&~~~{\cal E}(t)\equiv\displaystyle{ e^{-\frac{\mu}{12\sqrt{2}}\gamma^{789}\, t} {\cal E}}\,, \quad
\gamma^{24}{\cal E} = \gamma^{31}{\cal E} = \gamma^{789}{\cal E}\,, \quad {\cal E} = C{\cal E}^\ast \,,
\qquad p=5,6,\cdots,9\,. \label{dsusy}\eea
\\
From the dual relation, (\ref{dual}), one can also obtain the supersymmetry transformations of  the quadratic
D2 action, ${\cal S}^{\prime}$,  $(\ref{calSD2})$,
\be\ba{l} \delta \phi^l  = i \psi^\dagger \gamma^l {\cal E}(t)\,, \\
{}\\
\delta F_{\mu\nu} = i\partial_{\mu}[\psi^\dagger \gamma_{\nu}{\cal E}(t)]-
i\partial_{\nu}[\psi^\dagger \gamma_{\mu}{\cal E}(t)]+{\cal O}(\psi)\,,\\ {}\\
\delta \psi =  \left[\half F_{\mu\nu} \tilde{\gamma}^{\mu}\gamma^{\nu}  +  \partial_\mu \phi^l\tilde{\gamma}^{\mu}\gamma_l
-{\textstyle\frac{\mu}{12\sqrt{2}}} \phi^p(\gamma_p\gamma^{789}+3\gamma^{789}\gamma_p) \right] {\cal E}(t) \,,  \ea\ee
where $l=3,\cdots,9$,   $p=5,\cdots,9$, and ${\cal O}(\psi)$ denotes the terms which vanish when we impose
the equation of motion for $\psi$. Such terms are there since we integrated out  $\phi^{y}$ using its
equation of motion.  Nevertheless, the off-shell supersymmetry transformations of the  action, ${\cal
S}^{\prime}$,   are given by the above formulae without ${\cal O}(\psi)$ .  Finally, as done in the
subsection \ref{compare}, tuning  the gauge choices,  one can show that the above worldvolume supersymmetry
is consistent with  the one derived in the matrix model, (\ref{EEG}),  (\ref{WVsusytr}).

%%%%%%%%%%%%%%%%%%%%%%%%%%%%%%%%%%%%%%%%%%%%%%%%%%%%%%%%%%%%%%%%%%%%%%%%%%%%%%%%%%%%%%%%%%%%%%%%%%%%%%%%%%%%
\subsection{Supersymmetry algebra and the supermultiplets\label{sas}}
The supersymmetry algebra of the action, ${\cal S}$,  (\ref{M2action}), can be read off easily from our
previous work on the five dimensional theory \cite{HyunPark}, through the dimensional reduction.  The
supersymmetry algebra of the  3D  ${\N}=2$ worldvolume theory reads, with the Hamiltonian, $H$, $\so(2)$,
$\so(3)$ $R$-symmetry generators, $M_{56}$, $M_{rs}$, and real central charges, $\R$, $\R_{r}$, $\A_{r}$,
$\B_{r}$, $r=7,8,9$,
\begin{equation}
[H,Q]=0\,,\label{s1}
\end{equation}

\begin{equation}
\ba{ll} {}[M_{56},Q]=i\textstyle{\frac{1}{2}}\gamma_{56}Q\,,~~~~~&~~~~~
{}[M_{rs},Q]=i\textstyle{\frac{1}{2}}\gamma_{rs}Q\,,\\
{}&{}\\
{}[M_{r},M_{s}]=i\epsilon_{rst}M_{t}\,,~~~~~~~&~~~~~~M_{r}=\half\epsilon_{rst}M_{st}\,,\ea\label{s2}
\end{equation}

\begin{equation}
{}\{Q,\,Q^{\dagger}\}=2\Omega\Big[H-\R-\textstyle{\frac{\mu}{3\sqrt{2}}} M_{56}+\gamma^{r}(\R_{r}+\textstyle{\frac{\mu}{3\sqrt{2}}} M_{r})
+\gamma^{125r}\A_{r}+\gamma^{436r}\B_{r}\Big]\Omega\,.\label{susyalge}
\end{equation}
The supercharge is subject to
\begin{equation}
\ba{ll}Q=C(Q^{\dagger})^{T}\,,~~~~&~~~~Q=\Omega Q\,,\ea
\end{equation}
resulting in the four independent real components. The explicit forms of  $H$, $M_{56}$, $M_{rs}$, $\R$,
$\R_{r}$, $\A_{r}$, $\B_{r}$ are  given in the Appendix \ref{usefulB}. The numbers of degrees in the left
and right hand sides of (\ref{susyalge}) match as
\begin{equation}
10=1+3+3+3\,.
\end{equation}
Note that  $\Omega$, $\Omega\gamma^{r}\Omega$, $\Omega\gamma^{125r}\Omega$, $\Omega\gamma^{436r}\Omega$ are
the only allowed independent gamma matrix products to appear on the right. From the positive definity, we
have the following BPS energy bound,
\begin{equation}
H\geq \R+\textstyle{\frac{\mu}{3\sqrt{2}}}M_{56}+\left|(\hat{e}_{1})_{r}(\R_{r}+\textstyle{\frac{\mu}{3\sqrt{2}}}
M_{r})\right|+\left|(\hat{e}_{2})_{r}\A_{r}\right|+\left|(\hat{e}_{3})_{r}\B_{r}\right|\,,\label{Ebound}
\end{equation}
where $\hat{e}_{1}$, $\hat{e}_{2}$, $\hat{e}_{3}$ form an arbitrary orthonormal  real basis for the
``$7,8,9$'' space so that $(\hat{e}_{1})_{r}\gamma^{r}$, $(\hat{e}_{2})_{r}\gamma^{125r}$,
$(\hat{e}_{3})_{r}\gamma^{436r}$ can be simultaneously diagonalized with the
eigenvalues, $\pm 1$.\\

The energy spectra and the numbers of the corresponding  bosons and fermions  can be obtained by solving the
Abelian sector of the equations of motion,  (\ref{EOM}). They are summarized  in Table \ref{T1}.  Each row
forms an independent supermultiplet, and there exit three multiplets. Note that in three dimensions the gauge
fields have only one on-shell degree. In the present massive gauge theory, the nontrivial linear
combinations of the gauge fields and  the two Higgs, $\phi_{3}$, $\phi_{4}$,  form three independent
degrees, one for each multiplet.

\begin{table}[htb]
\begin{center}
\begin{tabular}{l|cccccc}
~~~energy spectra~~~~~~~~ & $~~~~\psi~~~~$ & $~A_{\mu},\phi_{3},\phi_{4}~$ &
$~~~\phi_{5},\phi_{6}~~~$ & $~~~\phi_{7},\phi_{8},\phi_{9}~~~$\\
 \hline
{}&{}&{}&{}\\
$E_{{\bf k}}=\sqrt{(\textstyle{\frac{\mu}{3}})^{2}+{{\bf
k}}^{2}\,}$ & 4 & 1 & 0 & 3\\
{}&{}&{}&{}\\
$E^{+}_{{\bf k}}=\sqrt{(\textstyle{\frac{\mu}{6}})^{2}+{{\bf
k}}^{2}\,}+\textstyle{\frac{|\mu|}{6}}$ & 2 & 1 & 1 & 0\\
{}&{}&{}&{}\\
$E^{-}_{{\bf k}}=\sqrt{(\textstyle{\frac{\mu}{6}})^{2}+{{\bf
k}}^{2}\,}-\textstyle{\frac{|\mu|}{6}}~~~$ & 2 & 1 & 1 & 0\\
%{}&{}&{}&{}\\
%\hline
\end{tabular}
\end{center}
\caption{Energy spectra and the numbers of bosons and fermions.} \label{T1}
\end{table}

%%%%%%%%%%%%%%%%%%%%%%%%%%%%%%%%%%%%%%%%%%%%%%%%%%%%%%%%%%%%%%%%%%%%%%%%%%%%%%%%%%%%%%%%%%%%%%%%%%%%%%%%%%%%
%%%%%%%%%%%%%%%%%%%%%%%%%%%%%%%%%%%%%%%%%%%%%%%%%%%%%%%%%%%%%%%%%%%%%%%%%%%%%%%%%%%%%%%%%%%%%%%%%%%%%%%%%%%%
\subsection{BPS equations for the fully supersymmetric  configurations and vacua\label{ssBPS}}
In this subsection, we consider the BPS equations which describe  the configurations  preserving    all the
four supersymmetries. In the conventional supersymmetric models, such fully supersymmetric configurations
would be vacua, but in the present case, the novel structure of the supersymmetry algebra allows nontrivial
fully supersymmetric BPS configurations. They have the energy saturation,
\begin{equation}
H=\R+\textstyle{\frac{\mu}{3\sqrt{2}}}M_{56}\,,
\end{equation}
while other central and $R$-symmetry charges  vanish, $\R_{r}=\A_{r}=\B_{r}=M_{r}=0$.\\

The corresponding BPS equations can be obtained either by writing
$H-\R-\textstyle{\frac{\mu}{3\sqrt{2}}}M_{56}$  as a sum of squares or from the supersymmetry transformation
of the fermions \cite{Park:2002cb,jhpBPS}. The BPS equations are
\begin{equation}
\ba{ll}
F_{0\mu}=D_{0}\phi_{l}=D_{0}\phi_{r}=0\,,~~~&~~~D_{0}\phi_{p}-\textstyle{\frac{\mu}{3\sqrt{2}}}\epsilon_{pq}\phi_{q}=0\,,\\
{}&{}\\
{}[\phi_{r},\phi_{s}]-i\textstyle{\frac{\mu}{3\sqrt{2}}}\epsilon_{rst}\phi_{t}=0\,,~~~&~~~
D_{j}\phi_{r}=[\phi_{l},\phi_{r}]=[\phi_{p},\phi_{r}]=0\,,\\
{}&{}\\
F_{12}-i[\phi_{3},\phi_{4}]=0\,,~~~&~~~D_{1}\phi_{3}-D_{2}\phi_{4}=0\,,\\
{}&{}\\
D_{1}\phi_{4}+D_{2}\phi_{3}-i[\phi_{5},\phi_{6}]=0\,,~~~&~~~D_{j}\phi_{p}+iJ_{jl}\epsilon_{pq}[\phi_{l},\phi_{q}]=0\,, \ea
\end{equation}
where $j=1,2$, $l=3,4$, $p=5,6$,  $r=7,8,9$, $J_{14}=J_{23}=\epsilon_{56}=1$.   The BPS equations themselves
satisfy the Gauss constraint so that any BPS solution satisfies the full equations of motion,  (\ref{EOM}).
The last four BPS equations are essentially the dimensional reduction of  the
BPS equations in the 6D Euclidean pure super Yang-Mills  \cite{jhpBPS}.\\

The classical supersymmetric vacua are given by the constant fuzzy spheres and arbitrary vevs for $\phi_{3}$,
$\phi_{4}$,
\begin{equation}
\ba{llll} [\phi_{r},\phi_{s}]=i\textstyle{\frac{\mu}{3\sqrt{2}}}\epsilon_{rst}\phi_{t}\,,~~~~&~~~
\phi_{3}=c_{3}\,,~~~~&~~~\phi_{4}=c_{4}\,,~~~~&~~~\phi_{5}=\phi_{6}=F_{\mu\nu}=0\,. \ea
\end{equation}
From (\ref{dual}),   after  tuning of the gauge choices, the dual relation between the field strength and
the compact  scalar becomes
\begin{equation}
\ba{ccc} F_{12}=-\partial_{0}\phi^{y}\,,~~~&~~~F_{20}=\partial_{1}\phi^{y}+\textstyle{\frac{\mu}{3\sqrt{2}}}\phi^{3}\,,
~~~&~~~F_{01}=\partial_{2}\phi^{y}-\textstyle{\frac{\mu}{3\sqrt{2}}}\phi^{4}\,. \ea\label{dual2}
\end{equation}
Therefore, geometrically viewed  from the   eleven dimensions, the vacua correspond to the giant  graviton
  plus  the  membranes tilted to the eleventh direction,
\begin{equation}
\phi^{y}=-\textstyle{\frac{\mu}{3\sqrt{2}}}c_{3}x^{1}+\textstyle{\frac{\mu}{3\sqrt{2}}}c_{4}x^{2}\,.\label{tilt}
\end{equation}

\acknowledgments{The work of SH and SY was supported in part by grant No. R01-2000-00021 from the Basic Research Program of the Korea Science and
Engineering Foundation.}
%%%%%%%%%%%%%%%%%%%%%%%%%%%%%%%%%%%%%%%%%%%%%%%%%%%%%%%%%%%%%%%%%%%%%%%%%%%%%%%%%%%%%%%%%%%%%%%%%%%%%%%%%%%%%%%
%%%%%%%%%%%%%%%%%%%%%%%%%%%%%%%%%%%%%%%%%%%%%%%%%%%%%%%%%%%%%%%%%%%%%%%%%%%%%%%%%%%%%%%%%%%%%%%%%%%%%%%%%%%%%%%
%%%%%%%%%%%%%%%%%%%%%%%%%%%%%%%%%%%%%%%%%%%%%%%%%%%%%%%%%%%%%%%%%%%%%%%%%%%%%%%%%%%%%%%%%%%%%%%%%%%%%%%%%%%%%%%
%%%%%%%%%%%%%%%%%%%%%%%%%%%%%%%%%%%%%%%%%%%%%%%%%%%%%%%%%%%%%%%%%%%%%%%%%%%%%%%%%%%%%%%%%%%%%%%%%%%%%%%%%%%%%%%
\newpage
\appendix

\begin{center}
\large{\textbf{Appendix}}
\end{center}
\setcounter{equation}{0}
\renewcommand{\theequation}{A.\arabic{equation}}

\section{Conventions and useful formulae \label{Appendix}}
\subsection{In  the supergravity setup\label{usefulA}}
To make a connection to the matrix model,  we choose the following representation of the flat eleven
dimensional spacetime  gamma matrices,
\be \Gamma^{\hat{S}} =\left(\begin{array}{cc} 0  & \tilde{\gamma}^{\hat{S}} \\
\gamma^{\hat{S}} & 0
\end{array}\right) \,,\qquad
\Gamma^{\hat{y}} =\Gamma^{\hat{0}\hat{1}\cdots \hat{9}}=\left(\begin{array}{rr} 1  & 0  \\
0 & -1
\end{array}\right) \,,\label{11Dgamma}
\ee
where $\hat{S}=0,1,\cdots,9$ and, in terms of the Euclidean nine dimensional gamma matrices, $\gamma^{A},\,A=1,2,\cdots,9$,
\be \tilde{\gamma}^{\hat{R}}=(1,\gamma^A)\,,\qquad \gamma^{\hat{R}}=(-1,\gamma^A)\,. \ee
Thus, in terms of the 9D Euclidean charge conjugate matrix, $C$, (\ref{chargeC}), the eleven dimensional
complex conjugate matrix, ${\cal B}$, is written as
\begin{equation}
\ba{cc} {\cal
  B}=\left(\ba{cc}C&0\\0&C\ea\right)\,,~~~&~~~\left(\Gamma^{\hat{R}}\right)^{\ast}={\cal B}^{-1}\Gamma^{\hat{R}}{\cal B}\,,~~~~~\hat{R}=0,1,\cdots,10\,.\ea
\end{equation}
The 32-component 11D Majorana spinor, $\theta$, satisfies
\begin{equation}
\theta={\cal B}\theta^{\ast}\,.\label{11DMajorana}
\end{equation}

We take the vielbein of the pp-wave metric, (\ref{metric}),  as follows,
\bea &&e^{\hat{t}} = (1+H/2)^{1/2}\Big(dt+\frac{H/2}{1+H/2}dy\Big)\,,\qquad
e^{\hat{y}} = (1+H/2)^{-1/2}dy \,, \nonumber \\
\nonumber \\
 &&e^{\hat{1}} = dx^1 + \textstyle{\frac{\mu}{3\sqrt{2}}}x^4
(dt+dy)\,, \qquad
e^{\hat{2}} = dx^2 + \textstyle{\frac{\mu}{3\sqrt{2}}x^3 (dt+dy)}\,,  \\
\nonumber \\
 &&e^{\hat{l}}= dx^l \,, \hskip2cm l=3,...,9\,, \nonumber\eea
from which the non-vanishing spin connections can be determined,
\be \ba{l} \omega_{\hat{t}\hat{y}\, l}
=-\frac{1}{4}(1+H/2)^{-1}\partial_{l}H\,,\qquad l=3,\cdots,9\,,\\
{}\\
\omega_{\hat{t}\hat{l}\, t}= \omega_{\hat{t}\hat{l}\, y} = \omega_{\hat{y}\hat{l}\, y}=\omega_{\hat{y}\hat{l}\, t}=
\frac{1}{4}(1+H/2)^{-1/2}\partial_lH - (\frac{\mu}{6})^{2}(1+H/2)^{-1/2}(x^3\delta_{\hat{l}\hat{3}}+x^4\delta_{\hat{l}\hat{4}}) \,, \\{}\\
\omega_{\hat{t}\hat{1}\, 4} = \omega_{\hat{t}\hat{4}\, 1} = \omega_{\hat{t}\hat{2}\, 3} =\omega_{\hat{t}\hat{3}\, 2}=\omega_{\hat{y}\hat{1}\, 4}=
\omega_{\hat{y}\hat{4}\, 1} = \omega_{\hat{y}\hat{2}\, 3} = \omega_{\hat{y}\hat{3}\, 2} = -\frac{\mu}{6\sqrt{2}}(1+H/2)^{-1/2} \,,\\
{}\\
\omega_{\hat{1}\hat{4}\, t} = \omega_{\hat{1}\hat{4}\, y} =
 \omega_{\hat{2}\hat{3}\, t}= \omega_{\hat{2}\hat{3}\, y}
 =-\frac{\mu}{6\sqrt{2}}\,.
\ea \ee
The explicit forms of the curved spacetime gamma matrices are
\be \ba{l} \Gamma_t = (1+H/2)^{1/2}\Gamma_{\hat{t}} + \frac{\mu}{3\sqrt{2}}x^4\Gamma_{\hat{1}} +
\frac{\mu}{3\sqrt{2}}x^3\Gamma_{\hat{2}}  \,, \\
{}\\
\Gamma_y = (1+H/2)^{-1/2}\Gamma_{\hat{y}} + \frac{H}{2}(1+H/2)^{-1/2}\Gamma_{\hat{t}} + \frac{\mu}{3\sqrt{2}}x^4\Gamma_{\hat{1}} +
\frac{\mu}{3\sqrt{2}}x^3\Gamma_{\hat{2}}  \,, \\
{}\\
\Gamma_{1} = \Gamma_{\hat{1}}\, , \qquad \Gamma_2=\Gamma_{\hat{2}}\, , \qquad \Gamma_l=\Gamma_{\hat{l}}\,,~~~l=3,\cdots,9\,.\ea\ee
In the given pp-wave background geometry, (\ref{metric}), the supersymmetry variations of the gravitino reduce to, with $\partial_\pm
\equiv\frac{1}{\sqrt{2}}(\partial_t\pm \partial_y)$, $n=1,\cdots, 9$,
\be\ba{l} \frac{1}{\sqrt{2}}(\delta\psi_t+\delta\psi_y)=\!\Big[\partial_{+}+\Omega_{+}-\frac{1}{4}\partial_nH(1+H/2)^{-1/2}\Gamma^{\hat{+}\hat{n}}
                 +\frac{\mu}{3}(1+H/2)^{-1/2}(x^3\Omega_2+x^4\Omega_1)\Big]\eta\,, \\
{}\\
\frac{1}{\sqrt{2}}(\delta\psi_t-\delta\psi_y) =\!\partial_{-}\eta \,,\qquad
 \delta\psi_n = \!\Big[ \partial_n +
\frac{1}{8}\partial_nH(1+H/2)^{-1}\Gamma^{\hat{t}\hat{y}} + (1+H/2)^{-1/2}\Omega_n \Big]\eta\,. \ea\ee

Some useful relations to derive the Killing spinors, (\ref{Killing}),   are
\bea &[\Omega_1, \Omega_+]&=[\Omega_2, \Omega_+]=0\,, \label{o12} \\
\nonumber \\
& [\Omega_l, \Omega_+] &= \left\{
\begin{array}{ll}
\frac{\mu}{3}\Omega_l\Gamma^{\hat{7}\hat{8}\hat{9}}\,,   &\qquad l=3,4 \,,
 \\ \\
\frac{\mu^2}{72}\Gamma^{\hat{+}\hat{l}}\,, & \qquad l=5,6 \,,
 \\ \\
\frac{\mu^2}{18}\Gamma^{\hat{+}\hat{l}}\,, & \qquad l=7,8,9 \,.
 \end{array} \right.\eea

\subsection{In the 3D $\N=2$ massive super Yang-Mills action\label{usefulB}}
The explicit forms of the supercharge, $Q$,  Hamiltonian, $H$, $\so(2)$, $\so(3)$ $R$-symmetry generators, $M_{56}$, $M_{rs}$, and real central
charges, $\R$, $\R_{r}$, $\A_{r}$, $\B_{r}$ are
\begin{equation}
\ba{l}Q=\Omega\displaystyle{\int {dx^{2}}\,\trN\Big[-\half
F_{\mu\nu}\gamma^{\mu}\tilde{\gamma}^{\nu}+D_{\mu}\phi_{a}\gamma^{a}\tilde{\gamma}^{\mu}+i\half
[\phi_{a},\phi_{b}]\gamma^{ab} +\textstyle{\frac{\mu}{3\sqrt{2}}}\phi_{a}\gamma^{a}\gamma^{789} \Big]\psi}\,,\\
{}\\
H=\displaystyle{\int dx^{2}\,\trN\Big[\half F_{0i}^{2}+\half F_{12}^{2}+\half D_{0}\phi_{a}^{2}+\half
D_{i}\phi_{a}^{2}-\quarter[\phi_{a},\phi_{b}]^{2}
+i\textstyle{\frac{\mu}{3\sqrt{2}}}\epsilon^{rst}\phi_{r}\phi_{s}\phi_{t}+\half(\textstyle{\frac{\mu}{3\sqrt{2}}})^{2}\phi_{r}^{2}\Big]\,,}\\
{}\\
M_{56}=\displaystyle{\int dx^{2}\,}\trN\Big[\epsilon^{pq}D_{0}\phi_{p}\phi_{q}-\textstyle{\frac{\mu}{6\sqrt{2}}}(\phi_{5}^{2}+
\phi_{6}^{2})-i\textstyle{\frac{1}{4}}\psi^{\dagger}\gamma_{56}\psi\Big]\,,\\
\\
M_{rs}=\displaystyle{\int
dx^{2}\,}\trN\Big[D_{0}\phi_{r}\phi_{s}-D_{0}\phi_{s}\phi_{r}-i\textstyle{\frac{1}{4}}\psi^{\dagger}\gamma_{rs}\psi\Big]\,,\\{} \ea
\end{equation}

\begin{equation}
\ba{l}
\R=\displaystyle{\int dx^{2}\,\partial_{i}\,\trN
\Big(\half\epsilon^{ij}(\phi_{3}D_{j}\phi_{4}-\phi_{4}D_{j}\phi_{3})-iJ^{il}\phi_{5}[\phi_{l},\phi_{6}]\Big)\,,}\\
{}\\
\R_{r}=-i\half\epsilon_{rst}\displaystyle{\int
dx^{2}\,\partial_{i}\,\trN\left(J^{il}[\phi_{l},\phi_{s}]\phi_{t}\right)\,,}\\
{}\\
\A_{r}=-\displaystyle{\int
dx^{2}\,\partial_{i}\,\epsilon^{ij}\trN\Big(\phi_{r}D_{j}\phi_{5}+iJ_{jl}[\phi^{l},\phi_{6}]\Big)\,,}\\
{}\\
\B_{r}=\displaystyle{\int dx^{2}\,\partial_{i}\,\epsilon^{ij}\trN\Big(\phi_{r}D_{j}\phi_{6}-iJ_{jl}[\phi^{l},\phi_{5}]\Big)\,,}\ea
\end{equation}
where $a=3,\cdots,9$, $i=1,2$, $l=3,4$, $r=7,8,9$ and  $\epsilon^{12}=J^{14}=J^{23}=J_{14}=J_{23}=1$.\\

The equations of motion are
\begin{equation}
\ba{l} D_{\nu}F^{\nu}{}_{0}+i[\phi_{a},D_{0}\phi_{a}] +\half\{\psi^{\dagger}{}^{\alpha},\psi_{\alpha}\}
-\textstyle{\frac{\mu}{3\sqrt{2}}}(D_{1}\phi_{4}+D_{2}\phi_{3}+i[\phi_{5},\phi_{6}])=0\,,\\
{}\\
D_{\nu}F^{\nu}{}_{i}+i[\phi_{a},D_{i}\phi_{a}] +\half\{\psi^{\dagger}{}^{\alpha},(\gamma_{i}\psi)_{\alpha}\}
-\textstyle{\frac{\mu}{3\sqrt{2}}}J_{il}D_{0}\phi_{l}=0\,,\\
{}\\
D_{\mu}D^{\mu}\phi_{l}-[\phi_{a},[\phi_{a},\phi_{l}]]
+\half\{\psi^{\dagger}{}^{\alpha},(\gamma_{l}\psi)_{\alpha}\}-\textstyle{\frac{\mu}{3\sqrt{2}}}J_{li}F_{0i}=0\,,\\
{}\\
D_{\mu}D^{\mu}\phi_{p}-[\phi_{a},[\phi_{a},\phi_{p}]]
+\half\{\psi^{\dagger}{}^{\alpha},(\gamma_{p}\psi)_{\alpha}\}+\textstyle{\frac{\mu}{3\sqrt{2}}}\epsilon_{pq}D_{0}\phi_{q}=0\,,\\
{}\\
D_{\mu}D^{\mu}\phi_{r}-[\phi_{a},[\phi_{a},\phi_{r}]] +\half\{\psi^{\dagger}{}^{\alpha},(\gamma_{r}\psi)_{\alpha}\}
-i\textstyle{\frac{\mu}{\sqrt{2}}}\epsilon_{rst}\phi_{s}\phi_{t}-(\textstyle{\frac{\mu}{3\sqrt{2}}})^{2}\phi_{r}=0\,,\\
{}\\
\gamma^{\mu}D_{\mu}\psi-i\gamma^{a}[\phi_{a},\psi] -\textstyle{\frac{\mu}{12\sqrt{2}}}(\gamma^{14}+\gamma^{23}-\gamma^{56}+3\gamma^{789})\psi=0\,,
\ea\label{EOM}
\end{equation}
where $i=1,2$, $l=3,4$, $p=5,6$, $r=7,8,9$ and  $J_{14}=J_{23}=1$.


\newpage


\begin{thebibliography}{99}


\bibitem{D-brane}
J.~Polchinski, S.~Chaudhuri and C.~V.~Johnson, ``Notes on D-Branes,'' [hep-th/9602052];\\
%%CITATION = HEP-TH 9602052;%%
J.~Polchinski,  ``Lectures on D-branes,'' [hep-th/9611050].
%%CITATION = HEP-TH 9611050;%%









\bibitem{Berenstein:2002jq}
D.~Berenstein, J.~M.~Maldacena and H.~Nastase,
%``Strings in flat space and pp waves from N = 4 super Yang Mills,''
JHEP {\bf 0204} (2002) 013 [hep-th/0202021].
%%CITATION = HEP-TH 0202021;%%



%\cite{Metsaev:2001bj}
\bibitem{Metsaev:2001bj}
R.~R.~Metsaev,
%``Type IIB Green-Schwarz superstring in plane wave Ramond-Ramond  background,''
Nucl.\ Phys.\ B {\bf 625} (2002) 70 [hep-th/0112044].
%%CITATION = HEP-TH 0112044;%%



%\cite{Metsaev:2002re}
\bibitem{Metsaev:2002re}
R.~R.~Metsaev and A.~A.~Tseytlin,
%``Exactly solvable model of superstring in plane wave Ramond-Ramond  background,''
Phys.\ Rev.\ D {\bf 65} (2002) 126004 [hep-th/0202109].
%%CITATION = HEP-TH 0202109;%%


%\cite{Brecher:2002ar}
\bibitem{Brecher:2002ar}
D.~Brecher, C.~V.~Johnson, K.~J.~Lovis and R.~C.~Myers,
%``Penrose limits, deformed pp-waves and the string duals of N = 1 large N  gauge theory,''
JHEP {\bf 0210} (2002) 008 [hep-th/0206045].
%%CITATION = HEP-TH 0206045;%%





%\cite{Alishahiha:2002nf}
\bibitem{Alishahiha:2002nf}
M.~Alishahiha, M.~A.~Ganjali, A.~Ghodsi and S.~Parvizi, ``On type IIA string theory on the PP-wave background,'' [hep-th/0207037].
%%CITATION = HEP-TH 0207037;%%





\bibitem{Hyun:2002wu}
S.~Hyun and H.~Shin,
%``N = (4,4) type IIA string theory on pp-wave background,''
JHEP {\bf 0210} (2002) 070 [hep-th/0208074].
%%CITATION = HEP-TH 0208074;%%


\bibitem{Hyun:2002wp}
S.~Hyun and H.~Shin, ``Solvable N = (4,4) type IIa string theory in plane-wave background and  D-branes,'' [hep-th/0210158].
%%CITATION = HEP-TH 0210158;%%



\bibitem{Kim:2002cr}
N.~Kim, K.~Lee and P.~Yi,
%``Deformed matrix theories with N = 8 and fivebranes in the pp wave  background,''
JHEP {\bf 0211} (2002) 009 [hep-th/0207264].
%%CITATION = HEP-TH 0207264;%%


\bibitem{Lee:2002vx}
K.~Lee,
%``M-theory on less supersymmetric pp-waves,''
Phys.\ Lett.\ B {\bf 549} (2002) 213 [hep-th/0209009].
%%CITATION = HEP-TH 0209009;%%




%\cite{HyunPark}
\bibitem{HyunPark}
S. Hyun and J.-H. Park,
%``5D action for longitudinal five branes on a pp-wave,''
JHEP {\bf 0211} (2002) 001 [hep-th/0209219].
%%CITATION = HEP-TH 0209219;%%









\bibitem{Metsaev:2002sg}
R.~R.~Metsaev, ``Supersymmetric D3 brane and N = 4 SYM actions in plane wave  backgrounds,'' [hep-th/0211178].
%%CITATION = HEP-TH 0211178;%%

%\cite{Metsaev:2003bf}
\bibitem{Metsaev:2003bf}
R.~R.~Metsaev, ``Superfield formulation of N=4 Super Yang-Mills theory in plane wave background,'' [hep-th/0301009].
%%CITATION = HEP-TH 0301009;%%

\bibitem{Bonelli:2002mb}
G.~Bonelli, ``Matrix strings in pp-wave backgrounds from deformed super Yang-Mills  theory,'' [hep-th/0205213].
%%CITATION = HEP-TH 0205213;%%




\bibitem{Banks:1996vh}
T.~Banks, W.~Fischler, S.~H.~Shenker and L.~Susskind,
%``M theory as a matrix model: A conjecture,''
Phys.\ Rev.\ D {\bf 55} (1997) 5112 [hep-th/9610043].
%%CITATION = HEP-TH 9610043;%%


\bibitem{Susskind:1997cw}
L.~Susskind, ``Another conjecture about M(atrix) theory,'' [hep-th/9704080].
%%CITATION = HEP-TH 9704080;%%


\bibitem{Sen}
A.~Sen,
%``D0 branes on T(n) and matrix theory,''
Adv.\ Theor.\ Math.\ Phys.\  {\bf 2} (1998) 51 [hep-th/9709220].
%%CITATION = HEP-TH 9709220;%%


\bibitem{Seiberg}
N.~Seiberg,
%``Why is the matrix model correct?,''
Phys.\ Rev.\ Lett.\  {\bf 79} (1997) 3577 [hep-th/9710009].
%%CITATION = HEP-TH 9710009;%%









\bibitem{KimPark}
N.~Kim and J.-H.~Park,
% ``Superalgebra for M-theory on a pp-wave,''
Phys.\ Rev.\ D {\bf 66} (2002) 106007 [hep-th/0207061].
%%CITATION = HEP-TH 0207061;%%



\bibitem{Keshav}
K. Dasgupta,  M. Sheikh-Jabbari and M. Van Raamsdonk, ``The BPS Spectrum of M-Theory on a PP-Wave,'' [hep-th/0207050].




\bibitem{Park:2002cb}
J.-H. Park,
%``Supersymmetric objects in the M-theory on a pp-wave,''
JHEP {\bf 0210} (2002) 032 [hep-th/0208161].
%%CITATION = HEP-TH 0208161;%%




\bibitem{Dasgupta:2002hx}
K.~Dasgupta, M.~M.~Sheikh-Jabbari and M.~Van Raamsdonk,
%``Matrix perturbation theory for M-theory on a PP-wave,''
JHEP {\bf 0205} (2002) 056 [hep-th/0205185].
%%CITATION = HEP-TH 0205185;%%



\bibitem{Kim:2002if}
N.~Kim and J.~Plefka, ``On the spectrum of pp-wave matrix theory,'' [hep-th/0207034].
%%CITATION = HEP-TH 0207034;%%


\bibitem{Maldacena:2002rb}
J.~Maldacena, M.~M.~Sheikh-Jabbari and M.~Van Raamsdonk,  ``Transverse fivebranes in matrix theory,'' [hep-th/0211139].
%%CITATION = HEP-TH 0211139;%%




\bibitem{Hyun:2002cm}
S.~Hyun and H.~Shin,
%``Branes from matrix theory in pp-wave background,''
Phys.\ Lett.\ B {\bf 543} (2002) 115 [hep-th/0206090].
%%CITATION = HEP-TH 0206090;%%



\bibitem{Bak:2002rq}
D.~Bak, ``Supersymmetric branes in PP wave background,'' [hep-th/0204033].
%%CITATION = HEP-TH 0204033;%%



\bibitem{Mikhailov:2002wx}
A.~Mikhailov, ``Nonspherical Giant Gravitons and Matrix Theory,'' [hep-th/0208077].
%%CITATION = HEP-TH 0208077;%%




\bibitem{NairKCS}
V.~P.~Nair, ``Kahler-Chern-Simons theory,'' [hep-th/9110042];\\
%%CITATION = HEP-TH 9110042;%%
V.~P.~Nair, ``Kahler-Chern-Simons theory,'' CU-TP-534 {\it Invited talks given at Strings and Symmetries '91, Stony Brook, N.Y., May 20-25, 1991 and
at 20th Int. Conf. on Differential Geometric Methods in Theoretical Physics, New York, N.Y., Jun
3-7, 1991};\\
V.~P.~Nair and J.~Schiff,
%``Kahler Chern-Simons theory and symmetries of antiselfdual gauge fields,''
Nucl.\ Phys.\ B {\bf 371} (1992) 329.
%%CITATION = NUPHA,B371,329;%%


\bibitem{Siegel:1979fr}
W.~Siegel,
%``Unextended Superfields In Extended Supersymmetry,''
Nucl.\ Phys.\ B {\bf 156} (1979) 135.
%%CITATION = NUPHA,B156,135;%%



%\cite{Skenderis:2002vf}
\bibitem{Skenderis:2002vf}
K.~Skenderis and M.~Taylor,
%``Branes in AdS and pp-wave spacetimes,''
JHEP {\bf 0206} (2002) 025 [hep-th/0204054].
%%CITATION = HEP-TH 0204054;%%




\bibitem{Kim:2002tj}
N.~Kim and J.~T.~Yee,  ``Supersymmetry and branes in M-theory plane-waves,'' [hep-th/0211029].
%%CITATION = HEP-TH 0211029;%%




\bibitem{Kowalski-Glikman}
J.~Kowalski-Glikman,
%``Vacuum States In Supersymmetric Kaluza-Klein Theory,''
Phys.\ Lett.\ B {\bf 134} (1984) 194.
%%CITATION = PHLTA,B134,194;%%

\bibitem{Figueroa-O'Farrill1}
J.~Figueroa-O'Farrill and G.~Papadopoulos,
%``Homogeneous fluxes, branes and a maximally supersymmetric solution of  M-theory,''
JHEP {\bf 0108} (2001) 036 [hep-th/0105308].
%%CITATION = HEP-TH 0105308;%%

\bibitem{FO2}
M.~Blau, J.~Figueroa-O'Farrill, C.~Hull and G.~Papadopoulos,
%``Penrose limits and maximal supersymmetry,''
Class.\ Quant.\ Grav.\  {\bf 19} (2002) L87 [hep-th/0201081].
%%CITATION = HEP-TH 0201081;%%






\bibitem{deWitpp}
B.~de Wit, K.~Peeters and J.~Plefka,
%``Superspace geometry for supermembrane backgrounds,''
Nucl.\ Phys.\ B {\bf 532} (1998) 99 [hep-th/9803209].
%%CITATION = HEP-TH 9803209;%%




\bibitem{deWit:1998yu}
B.~de Wit, K.~Peeters, J.~Plefka and A.~Sevrin,
%``The M-theory two-brane in AdS(4) x S(7) and AdS(7) x S(4),''
Phys.\ Lett.\ B {\bf 443} (1998) 153 [hep-th/9808052].
%%CITATION = HEP-TH 9808052;%%





\bibitem{JHPcomments}
D.~Bak, K.~Lee and J.-H.~Park,
%``Comments on noncommutative gauge theories,''
Phys.\ Lett.\ B {\bf 501} (2001) 305 [hep-th/0011244].
%%CITATION = HEP-TH 0011244;%%





\bibitem{jhpBPS}
D.~Bak, K.~Lee and J.-H.~Park, Phys. Rev. D {\bf 66} (2002) 025021 [hep-th/0204221].
%%CITATION = HEP-TH 0204221;%%




\bibitem{Bergshoeff:1997kr}
E.~Bergshoeff, R.~Kallosh, T.~Ortin and G.~Papadopoulos,
%``kappa-symmetry, supersymmetry and intersecting branes,''
Nucl.\ Phys.\ B {\bf 502} (1997) 149 [hep-th/9705040].
%%CITATION = HEP-TH 9705040;%%




\bibitem{Seiberg:2000zk}
N.~Seiberg,
%``A note on background independence in noncommutative gauge theories,  matrix model and tachyon condensation,''
JHEP {\bf 0009} (2000) 003 [hep-th/0008013].
%%CITATION = HEP-TH 0008013;%%




\bibitem{NC-CS}
D.~Bak, K.~M.~Lee and J.-H.~Park,
%``Chern-Simons theories on noncommutative plane,''
Phys.\ Rev.\ Lett.\  {\bf 87} (2001) 030402 [hep-th/0102188];\\
%%CITATION = HEP-TH 0102188;%%
V.~P.~Nair and A.~P.~Polychronakos,
%``On level quantization for the noncommutative Chern-Simons theory,''
Phys.\ Rev.\ Lett.\  {\bf 87} (2001) 030403 [hep-th/0102181];\\
%%CITATION = HEP-TH 0102181;%%
J.-H.~Park,
%``On a matrix model of level structure,''
Class.\ Quant.\ Grav.\  {\bf 19} (2002) L11 [hep-th/0108145].
%%CITATION = HEP-TH 0108145;%%




\bibitem{Cvetic:2002nh}
M.~Cvetic, H.~Lu, C.~N.~Pope and K.~S.~Stelle, ``Linearly-realised worldsheet supersymmetry in pp-wave background,'' [hep-th/0209193].
%%CITATION = HEP-TH 0209193;%%









\bibitem{Bergshoeff:1996tu}
E.~Bergshoeff and P.~K.~Townsend,
%``Super D-branes,''
Nucl.\ Phys.\ B {\bf 490} (1997) 145 [hep-th/9611173].
%%CITATION = HEP-TH 9611173;%%


\bibitem{Aganagic:1997zk}
M.~Aganagic, J.~Park, C.~Popescu and J.~H.~Schwarz,
%``Dual D-brane actions,''
Nucl.\ Phys.\ B {\bf 496} (1997) 215 [hep-th/9702133].
%%CITATION = HEP-TH 9702133;%%





%\cite{Michelson:2002wa}
\bibitem{Michelson:2002wa}
J.~Michelson,
%``(Twisted) toroidal compactification of pp-waves,''
Phys.\ Rev.\ D {\bf 66} (2002) 066002 [hep-th/0203140].
%%CITATION = HEP-TH 0203140;%%



\bibitem{Cederwall:1996ri}
M.~Cederwall, A.~von Gussich, B.~E.~Nilsson, P.~Sundell and A.~Westerberg,
%``The Dirichlet super-p-branes in ten-dimensional type IIA and IIB  supergravity,''
Nucl.\ Phys.\ B {\bf 490} (1997) 179 [hep-th/9611159].
%%CITATION = HEP-TH 9611159;%%





\bibitem{NDBI}
T. Hagiwara, J. Phys. A {\bf 14} (1981) 3059;\\
P.~C. Argyres and C.~R. Nappi, Nucl. Phys. B {\bf 330} (1990) 151;\\
A.~A.~Tseytlin,
%``On non-abelian generalisation of the Born-Infeld action in string  theory,''
Nucl.\ Phys.\ B {\bf 501} (1997) 41 [hep-th/9701125];\\
%%CITATION = HEP-TH 9701125;%%
J.-H. Park,
%``A study of a non-Abelian generalization of the Born-Infeld action,''
Phys.\ Lett.\ B {\bf 458} (1999) 471 [hep-th/9902081].
%%CITATION = HEP-TH 9902081;%%





\bibitem{Myers:1999ps}
R.~C.~Myers,
%``Dielectric-branes,''
JHEP {\bf 9912} (1999) 022 [hep-th/9910053].
%%CITATION = HEP-TH 9910053;%%











\end{thebibliography}
\end{document}